\documentclass[12pt,reqno]{amsart}
\usepackage{amsmath,amscd,amsfonts}

\newfont{\eu}{eusb10 at 12pt}

\newcommand{\bfcM}{{\boldsymbol{C}_0^1}}
\newcommand{\bfcm}{{\boldsymbol{c}_0^1}}
\newcommand{\hf}{{\hat f}}
\newcommand{\hx}{{\hat x}}
\newcommand{\tom}{T_{p_0}\m}
\newcommand{\omi}[2]{\omega_{#1#2}}
\newcommand{\oms}[2]{\omega^{#1#2}}

\newcommand{\gam}[3]{\gamma^#1_{#2#3}}
\newcommand{\diff}[1]{Dif\!f{}_{\mathrm{loc.}}^{\mathrm{#1}}}
\newcommand{\e}{\mbox{e}}
\newcommand{\Rset}{\mathbb{R}}
\newcommand{\sg}{{\sigma}}
\newcommand{\GammaG}{\Gamma_{\mathrm{\widehat{G}}}}
\newcommand{\SGammaH}{\Gamma_{\mathrm{\widehat{H}}}}

\newcommand{\m}{{\mathcal{M}}}
\newcommand{\Sub}{{\mathcal{S}}}
\newcommand{\Su}{{\mathcal{S}^1_c}}
\newcommand{\A}{{\mathcal{A}}}
\newcommand{\B}{{\mathcal{B}}}
\newcommand{\F}{{\mathcal{F}}}
\newcommand{\G}{{\mathcal{G}}}
\newcommand{\I}{{\mathcal{I}}}
\newcommand{\J}{{\mathcal{J}}}

\newcommand{\Lag}{{\mathcal{L}}}
\newcommand{\Sz}{{\mathcal{S}^0_c}}
\newcommand{\tsms}{\underline{T^\ast\Rset^n}}
\newcommand{\Oc}{\widehat{\Omega}}

\newcommand{\kc}{\hat{\kappa}}
\newcommand{\bfc}{{\pmb{c}}}
\newcommand{\bfah}{{\pmb{\hat{a}}}}
\newcommand{\bfa}{{\pmb{a}}}
\newcommand{\sgt}{\tilde{\sigma}}

\newcommand{\bftau}{\pmb{\tau}}
\renewcommand{\bysame}{\leavevmode\hbox
to3em{\hrulefill}\thinspace}
{\par\addvspace{6pt}\normalfont \normalsize}

{\theoremstyle{plain}%
  \newtheorem{theorem}{Theorem}

}                    

{\theoremstyle{remark}

}

{\theoremstyle{definition}
  \newtheorem{definition}{Definition}
  
}

\begin{document}


\title{Navigation of Spacetime Ships  in  Unified Gravitational
and Electromagnetic Waves} 
\author{
Jacques L. Rubin\\
\emph{Institut du Non-Lin{\'e}aire de Nice,\\
U.M.R. 6618 C.N.R.S. - Universit{\'e} de Nice - Sophia-Antipolis,\\
1361 route des Lucioles, 06560 Valbonne, France}\\
E-mail: jacques.rubin@inln.cnrs.fr} 
\date{Preprint INLN 2002/13, \today, v.3}
\footnote{This is the second version (v.3) of the
gr-qc/0205012 paper. Relative to the  second version,
some changes in the mathematical results have been 
done without consequences on the physical model (see also
hep-th/0404186). The conformally flatness of the substratum
spacetime, i.e.  the vanishing of the Weyl tensor associated
to the metric
$\underline{\omega}$, which  is an assumption used throughout
in  the mathematical developments from chapter 2, has  been
well precised in the first chapter. Clearer explanations at
the very end of  chapter 3 about accelerating frames are
given. New references are  indicated and some of them
corrected.}
\begin{abstract} 
On the basis of a \emph{``local"\/}  principle of equivalence of 
general relativity, we consider a navigation in a kind of
``4D-ocean" involving measurements of conformally invariant physical
properties only. Then, applying the Pfaff theory for PDE to a particular
conformally equivariant system of differential equations, we show the
dependency of any kind of function describing  ``spacetime waves", with
respect to 20 parametrizing functions. These latter, appearing in a
linear differential Spencer sequence and determining   gauge
fields of  deformations  relatively to  ``ship-metrics" or to 
``flat  spacetime ocean metrics", may be ascribed to unified electromagnetic 
and gravitational waves. The
present model is based neither on a classical gauge theory of 
gravitation or a gravitation theory with torsion, nor on any Kaluza-Klein
or Weyl type unifications, but  rather on a  post-Newtonian
approach of  gravitation in a four dimensional conformal
Cosserat spacetime.
\end{abstract}
\maketitle 
\bigskip
\par\bigskip
\par\bigskip

\vfill\eject

\section{Introduction: spacetime as a 4D-ocean}
\subsection{Smooth and striated spaces} 
Contrarily to what is usually known about the ether theory, A.
Einstein wasn't opposed to this very concept, but rather to the
concept of a favored frame and to a concept of ether considered as a
rigid object necessary to the propagation of light. After rejecting
the ether, he finally accepted it on the basis of the following
assumptions given in \emph{``Aether und
Relativitaetstheorie"\/} (Julius Springer Ed., Berlin, 1920, 15 pages.
From a lecture held on 5th may 1920 at Leiden University):
\begin{quotation}
p. 8:\emph{``Der n\"achstliegende Standpunkt, den man dieser Sachlage
gegen\-\"uber einnehmen konnte, schien der folgende zu sein. Der \"Ather
existiert \"uber\-haupt nicht."\/} (The most obvious viewpoint which
could be taken of this matter appeared to be the following. The ether
does not exist at all.)
\smallskip

p. 9:\emph{``Indessen lehrt ein genaueres Nachdenken, da\ss\, diese
Leugnung des \"Athers nicht notwendig durch das spezielle
Relativit\"ats\-prinzip gefordert wird. Man kann die Existenz eines
\"Athers annehmen; nur mu\ss\, man darauf verzichten, ihm einen
bestimmten Bewegungszustand zuzu\-schreiben, d. \!h. man mu\ss\, ihm
durch Abstraktion das letzte mechanische Merkmal nehmen, welches ihm
Lorentz noch gelassen hatte."\/}
(However, closer reflection shows that this denial of the ether
is not demanded by the special principle of relativity. We can assume
the existence of an ether, but we must abstain from ascribing a
definitive state of motion to it, i.e. we must by
abstraction divest it of the last mechanical characteristic which
Lorentz had left it.)
\smallskip

p. 10:\emph{``Verallgemeinernd m\"ussen wir sagen. Es lassen sich
ausgedehnte physikalische Gegenst\"ande denken, auf welche der
Bewegungsbegriff keine Anwendung finden kann\dots.
Das spe\-zielle Relativit\"atsprinzip verbietet uns, den \"Ather als
aus zeitlich verfolgbaren Teilchen bestehend anzuneh\-men, aber die
\"Ather\-hypo\-these an sich widerstreitet der spe\-ziellen
Relativit\"ats\-theorie nicht. Nur mu\ss\, man sich davor h\"uten, dem
\"Ather einen Bewegungszustand zuzusprechen."\/} (Generalizing, we must
say that we can conceive of extended physical objects to which the
concept of motion cannot be applied\dots. The special principle
of relativity forbids us to regard the ether as composed of particles,
the movements of which can be followed out through time, but the ether
hypothesis as such is not incompatible with the special theory of
relativity. Only we must take care not to ascribe a state of motion of
the ether.)
\end{quotation}
Moreover,  denying ether amounts to consider spacetime as deprived of any
physical properties, which is obviously not the case. This would suggest
a concept of ether as a medium referring to physical properties and not
to geometric or mechanical considerations (see p. \!\!12 in the reference
above). Somehow adopting this point of view, we will consider ether
as a {\em ``{\rm 4D}-ocean\/}" and later on, as a further specification,
as a {\em ``spacetime ocean"\/}. We recall that this analogy has been
made few years ago by W. G. Unruh \cite{unruhfluid}, and  detailed
discussions of the concepts involved in ether theory can be found in
\cite{kostrobook,levybook}.\par At first, in order to introduce as
clearly as possible our model of unification of electromagnetic and
gravitational forces, we will rely on metaphors in two and three
dimensions.\par\smallskip  Suppose a child asks  you: {\em ``At which
distance the blue sky is ?"\/}. Or likewise on a ``sailboat" looking
afar at a ship on a quiet blue  ocean:  {\em ``How far is that  ship~?"\/}. 
May be in a future orbital station, another child will wonder:
{\em `` Why can't we touch the stars with our hands ?"\/}.\par All
of these apparently naive questions send us back to one only but major
difficulty: the ocean, the sky, the outer space may be topological,
still surely non-metrical spaces (!), taking ``metrical" 
in its usual acceptation. In the case of a
spacetime ocean, assumed to be a non-metrical space, the latter remark 
forbids to conceive any kind of state of spacetime motions, since  a
notion of distance  would be required to evaluate, for instance, the
spacetime velocity. We know how difficult it is to evaluate distances on
sea, or altitudes of aircraft   just by looking at them (or evaluate
times without watches in spacetime). These kinds of spaces are not
{\em ``striated"\/} ones as our highways with permanent blank dashed
lines on ground allowing for evaluations of distances: the former are not
``naturally" endowed with fields of metrics. The coordinate maps, from
these spaces to $\Rset$-vector spaces, are only defined  on those
subspaces of points at which serial physical measurement processes are
performed (i.e. attributions of finite sets of numbers to some points or
some composed parts).
Distances could  be  undefined on those subspaces or, at best, only
defined on them. In fact, we might  have only a metric attached to each
point of a line,  a trajectory of a ship, the wake of  an airplane~\dots
and this metric would be built out of a local moving frame. The latter
could be the wings arrow of the airplane, the mast and the boom of a
sailboat, but also their wakes, furnishing a velocity vector or a flux
vector.  In fact, spacetime has a definite number of geometric
dimensions but the latter cannot be attributed  {\em ``a priori"\/} to
space or time dimensions: this ascription of an orientation can be
performed  only  {\em locally\/} in the moving frames.\par\smallskip 
Angle measurements only can be achieved with a calliper on a kind of
``sea horizon circle", or with a sextant on the well-known ``celestial
sphere": Now these are just the elements of a conformal geometry. Of
course these operations are strongly related to our eyesight and the
light, and it has been demonstrated in  two  classical 1910
studies of H. Bateman \cite{bateman} and E. Cunningham
\cite{cunningham} that the Maxwell equations of electromagnetic fields
in vacuum are  conformally equivariant (see also \cite{fulton62}). In sea
or aerial navigation, from  known beacons or hertzian markers out of
which angles can be measured, we can deduce the geometric positions on
charts. But, if just after, we lose the signals for a while (because of
fog or solar storms hiding the coast or perturbating electromagnetic
signals from markers), then we need to redirect our way in order to
recover them. But how this can be done without the use of magnetic
compasses or gyroscopes to orient  angles, and then to know the north,
the south, the bottom and the top ?\par Clearly we need magnetic and
gravitational forces to orient the moving frames or the angles, and this
means that, to be known, local geometry requires that forces be there
(!), to provide at least directions. These 2D and 3D spaces must be
endowed with a (magnetic or gravitational) field of orientation. On that
point, extending this approach to a 4D-ocean, we should have  such a
force to discriminate between  past and  future (which doesn't  mean  to
have an universal duration), allowing for instance to orient the light
cones. In some way, we would have to consider a ``time" force vector
``dressing" the ``spacetime", and assume a field
of time orientation added to the spacetime structure. We will come back
on that point in the conclusion since it is really a major historic
difficult question.\par It follows local geometries are deduced from 
forces and not the converse as is usually done in general relativity,
which, in this respect, appears to be in conflict with the principles of
navigations. We take the navigation side and the ``Prima" of the forces
on the geometry. Nevertheless, against appearances, the
\emph{``local"\/} principle of equivalence of  general relativity will be
kept throughout, but used  another way.\par  Here we present two
kinds of ``spaces" to which will be given synonyms depending on the context
(metaphorical, mathematical or physical):
\begin{itemize}
\item a {\em ``smooth
space"\/} that will be sometimes referred to as the {\em ``unfolded
spacetime"\/}, the
\mbox{\em ``\,{\rm 4D}-ocean"\/} or the {\em``spacetime ocean"\/}
and denoted by $\m$, and
\item a {\em ``striated space"\/} also dubbed the   {\em
``underlying or substratum spacetime"\/} or the {\em ``\,{\rm
4D}-ocean ground floor $\Sub$"\/}.
\end{itemize}
Our {\em ``ships", ``sailboats", ``aircraft"\/} would
be the tangent spaces
$T_{p_0}\m$ or moving frames, also viewed as the space of rulers,
callipers and watches. As a matter of fact, this model will involve an
unfolding or a deployment of the {\em ``smooth space"\/}
$\m$ from the {\em ``striated space"\/} $\Sub$ (we use the G. Deleuze
and F. Guattari terminology \cite[\S12 and \S14]{deleuzeguattari}). It
may also be viewed as a kind of generalisation in the PDE framework, of
the universal deployment concept for ODE introduced in particular by the
mathematician Ren\'e Thom in his well-known catastrophes theory. The
relevant modern mathematical terminology would be: cobordism theory.


\subsection{To tie a spacetime ship with its
environnement: the principle of equivalence}
Let us assume the unfolded spacetime $\m$  to be of class
$C^\infty$, of dimension $n\geq 4$, connected and paracompact. Let $p_0$
be a particular point in
$\m$, $U(p_0)$ an open neighborhood of $p_0$ in $\m$, and $\tom$ its
tangent space. The \emph{``local"\/} 
principle of equivalence we use (compatible with the usual ones. See a 
review of those principles in
\cite{ghins}), states it exists a local diffeomorphism
$\varphi_{p_0}$ attached to $p_0$ putting in a one-to-one
correspondence the points
$p\in U(p_0)$ with some vectors $\xi\in\tom$ in an open neighborhood of
the origin of $\tom$:
\[
\varphi_{p_0}:p\in
U(p_0)\subset\m\longrightarrow\xi\in\tom\,,\qquad
\varphi_{p_0}(p_0)=0\,.
\]
It is important to
remark  the correspondence  between ``position points" and
``position vectors". We will show in the next chapter it involves the
Einstein's principle of equivalence between relative uniformly accelerated
frames. To each position vector $\xi$ of $\tom$, we can associate a
frame made out of a ``little" local web of  straight lines
and thus we construct a {\it local\/} field of metrics
$\bar{g}_{p_0}(\xi)$ depending both on $p_0$ and the point $\xi$. This
field of metrics is defined on the tangent spaces of the tangent space:
$T_\xi(T_{p_0}\m)$. We will further assume that these  webs
are (partially) oriented with respect to the {\it local} orientation
provided by a time arrow (like the needle of a magnetic compass in 2D).
Hence a particular direction is selected among the four, and reflected in
the signature of the metric field 
$\bar{g}_{p_0}$ assumed to be of the Minkowski type $(+\,-\,-\,-)$ as
``usual".\par To proceed further, let us evoke a metaphor borrowed to the
Quattrocento painters: considering landscapes ($\Sub$), they wished to
draw them on canvas ($T_{p_0}\m$), seeing them through perspective grids
or webs ($\m$). In a 4D situation the ``grids" would deform themselves
and without an absolute grid of reference we may think instead of a
dynamical deformation of the landscapes.\par More precisely, we
consider other ``little" webs (at $p\in{}U(p_0)\subset\m$) on the
``surface" of the 4D-ocean $\m$. If we try to superpose  the latter with
those at $\xi\in\tom$ so as to simultaneously see their ``images",  we
make a projective conformal correspondence, as  painters did with their
geometric perspectives. Moreover if a  {\it local} metric
$\tilde{g}$  depending only on
$p\in\m$ is attached to this latter frame web, this correspondence means
that the metric $\bar{g}_{p_0}$ is pulled back by the application
$\varphi_{p_0}^\ast$ to  $\tilde{g}$. Then we have the relation: 
\(
\varphi_{p_0}^{\ast}(\bar{g}_{p_0})=\tilde{g}\,
\), and the metrics $\tilde{\omega}$ and
$\bar{\omega}$ on $\Rset^n$ of type $(+\,-\,-\,-)$
associated respectively to $\tilde{g}$ and
$\bar{g}_{p_0}$ should be thought of as being somehow proportional
in view of the previous metaphorical perspective. In other words they are
conformally equivariant with respect to {\it local\/} diffeomorphisms
$\hf_0$ of
$\Rset^n$ depending on $p_0$ and associated to $\varphi_{p_0}$, i.e.
conformally equivariant with respect  to the conformal Lie
pseudogroup associated with a metric field $\mu$ defined on
$\Rset^n$, of type $(+\,-\,-\,-)$. Thus we have in this case:
\begin{equation}
\hf^{\ast}_0(\bar{\omega})=\tilde{\omega}\,,
\qquad\tilde{\omega}\simeq_{p_0}e^{2\lambda_0}
\,\bar{\omega}\equiv\nu\,,\qquad\hf_0(0)=0\,,
\qquad\lambda_0(0)=0\,,
\label{tilom}
\end{equation}
and
\begin{equation}
\hf^{\ast}_0(\mu)=e^{2\alpha_0}\,\mu\,,\qquad\alpha_0(0)=0\,,
\label{mueq}
\end{equation}
where {\em $\lambda_0$ and $\alpha_0$ are  {\it local} functions
associated to each $\hf_0$\/} and also depending on $p_0$. Also we
indicate with the sign\, ``\,$\simeq_{p_0}$\," the {\it local\/}
equality at or for a given point $p_0$. In this expression we point out that
{\em $\lambda_0$ and $\alpha_0$ are functions\/} in full
generality. {\em It is an essential feature of the model under
consideration}. In case of constants, then 15 parameters would define
$\hf_{p_0}$ and it is known that nothing could be ascribed to
electromagnetism but only to a uniform gravitational field.
\par\smallskip
The metric $\tilde{g}$ emerges also from an other outlook but in the
same framework. Indeed we can consider the {\em
``susbstratum spacetime $\Sub$"\/} as a striated space, that is, a
metrical space (a metaphorical description might consist in referring to
an ocean ground floor made out of \textit{striated sand\/}). Thus, we
assume  it can be endowed with a {\it global\/} metric we can denote by
$\underline{g}$. Also we make the assumption that
$\Sub$ has a constant Riemaniann  curvature tensor and then is
``conformally flat", i.e. the  Weyl tensor is vanishing. This will ensure
the integrability of the conformal Lie pseudogroup from the H. Weyl
theorem \cite{weylkonf}.  In other words, the webs
``drawn" on
$\Sub$ are only transformed into each others,  at the same ground
level, with no changes in the
``perspective": no dilatation transformations occur. Hence this 
``substratum spacetime $\Sub$" is equivariant with respect to
{\it global\/} diffeomorphisms $\psi$ of the Poincar\'e Lie
pseudogroup associated to a {\it global} metric field denoted
$\underline{g}$ and satisfying, for every {\it global\/}
diffeomorphism $\psi$ of $\Sub$,
\(
\psi^{\ast}(\underline{g})=\underline{g}\,
\).
If $\underline{\omega}$ represents $\underline{g}$ on $\Rset^n$, and if the
global diffeomorphims $\psi$ are associated to  
global applications $f$ of the Poincar\'e Lie pseudogroup of
$\Rset^n$, endowed again with the metric field $\mu$, then we have
\begin{equation}
f^{\ast}(\underline{\omega})=\underline{\omega}\,.
\label{omep}
\end{equation}
Continuing with the metaphorical description, 
we try to superpose again the web  at
$p\in\m$, and a  web ``drawn" on $\Sub$.
Each one is  just a mathematically \emph{smooth\/} but, a priori at that
step, a completly general  deformation  of the other. As a  consequence,
it exists {\it local\/} diffeomorphisms
$\phi_p$ such that 
\(
\phi^{\ast}_p(\underline{g})=\tilde{g}\,
\).
Then from the latter and expression \eqref{tilom}, we have on
$\Rset^n$, the relation
\begin{equation}
h^{\ast}_p(\underline{\omega})=\bar{\omega}\,,\qquad
h_p(0)=0\,,
\label{hdef}
\end{equation}
where $h_p$ is a {\it local\/} diffeomorphism of $\Rset^n$.
{\em Then we make the simplification assumption:
the metric field $\mu$ is identified with
$\underline{\omega}$, so that one obtains from {\eqref{mueq}} the
relation:}
\begin{equation}
\hf^{\ast}_0(\underline{\omega})=e^{2\alpha_0}\underline{\omega}\,.
\label{first}
\end{equation}
But also from  relations \eqref{tilom} and
\eqref{hdef}, we  can deduce the following commutative diagram:
\[
\begin{CD}
\displaystyle\underline{\omega}@>\displaystyle
f^\ast>>\underline{\omega}\\
@V\displaystyle{h_p^\ast}VV@VV\displaystyle{e^{2\,\lambda_0}\,h_p^\ast}V\\
\bar{\omega}@>>\displaystyle\hf_0^\ast>\nu
\end{CD}
\]
Therefore, to pass from $\underline{\omega}$ to $\bar{\omega}$ (or
from a point $\underline{p}\in\Sub$ to a ``point of deformation" $\xi\in
T_{p_0}\m$) is equivalent to pass from
$f$ to
$\hf_0$.\par Unfortunately (!), all the measurements are performed at
$p_0$. That means we  only consider angles  as well as the metric field
$\bar{\omega}$ up to any multiplicative factor, i.e. we must
consider the metric on the projective 3-dimensional 
space attached to $p_0$, namely
${\bf P}\Rset^{n-1}$. Hence  only the diffeomorphisms
$\hf_{0}$ giving any metric field
$\nu$ different from $\bar{\omega}$ (or equivalently
$\underline{\omega}$) will matter for physical applications as we
will see later on.\par  {\em To summarize, to pass from the substratum
spacetime
$\Sub$ with the metric field $\underline{\omega}$ to the tangent
spacetime
$T_{p_0}\m$ with the metric field $\bar{\omega}$ is equivalent to
pass from the Poincar\'e Lie pseudogroup to the conformal one, the
two being associated with~$\underline{\omega}$. Hence this
deformation process is associated to the coset of the two previous
pseudogroups. It is a set of functions parametrizing all the
deformation diffeomorphisms\, $h$. These functions are strongly
related to the arbitrary functions family $\alpha_0$ and $\lambda_0$, as
well as their derivatives. And we will show that they can be
related to the electromagnetic and gravitational gauge
potentials in a  unified way.}\par\medskip Of course, this very
classical approach in deformation theory differs from the classic gauge
one in general relativity  (see a review for instance in
\cite{ivan}). Indeed  the latter are developed from a
given gauge Lie group, which can be  Poincar\'e, 
conformal, or almost any other Lie group. But first, they are not
pseudogroups, and in second place  they are assossiated to Lie groups 
invariance of the tangent spaces (not the tangent fiber bundle) at any
fixed base point $p_0$. In fact that means  they are isotropy Lie
subgroups of the corresponding pseudogroups, or equivalently Lie groups
of the fibers of the tangent  bundles, namely structural Lie
groups. For instance, in fixing the function
$\alpha_0$ to a constant, meaning fixing $p_0$, the set of
applications $\hf_0$ becomes a Lie group and not a Lie pseudogroup. In
that case the applications $\hf_0$ would depend only on 15 real
variables (if $n=4$), not on a  set of arbitrary functions as we will see
for the conformal pseudogroup.\par\medskip 
Also it is not a Kaluza-Klein type
theory since  we just consider a spacetime with dimension $n=4$,  even
though the results, we present here, can be extended to higher
dimensions $n$. The fields of interactions don't come from the
addition of extra geometrical dimensions other than those necessary to
the geometrical   description of our four dimensional physical
spacetime.\par\medskip 
Sometimes  an unification is suggested in
considering a model based on a possible torsion of an affine connection,
i.e. based on a Riemann-Cartan geometry. In the present model, the
torsion is merely associated to motions of  ships, aircraft, cars such as
pitching, rolling, precession, rotation, etc\dots. The skew-symmetric
part of the connection symbols such as the Riemann-Christoffel ones (no
torsion) is related to these latter kinds of motions and not to  a sort
of Faraday tensor. In fact these  motions are only associated to  kinds
of constraints (such an example is provided by contraints of Lorentz
invariance in the Thomas precession mechanism) between a vector
$\eta$ and its tangent space and  not  to the base space $\m$, i.e. the
environment with its ``4D-waves". 
\par\medskip Also it differs completly from H. Weyl
unifications 	and the J.-M. Souriau approach \cite{souriau}. In fact
there were two different theories proposed by H. Weyl, in 1918  and
latter on in 1928-29
\cite{weylOfar,weylbook}. These theories  were based on
variational equivariance with respect to local \emph{given\/}
dilatation transformations, exhibiting some kinds of  Noether
invariants ascribed to electric charges or electric currents. In some
way, the model we present is  also related to dilatation transformations
but not to a variational problem and/or to topological invariants.
Moreover, our model deals  precisely with the determination of  space
of dilatation operations that are compatible with the conformal
transformations preserving the equivariance a given metric field. As
should become clear in next chapter 3, not all dilatations are
permitted, since they are bounded by constraints reflected in particular
functional dependencies. The latter will reflect themselves through the
existence of parametrizing functions which will be physically  ascribed
to fields or gauge potentials of interactions or of
spacetime deformations.
\par\smallskip 
Close approaches to ours, are
developed on the one hand by M. O. Katanaev \& I. V. Volovich
\cite{katavol92}, and on the other hand by H. Kleinert \cite{kleinbook2}
and J.-F. Pommaret \cite{pommbook94}. These authors, however, do not seem
to have been  concerned with any model of unification, since a
particular system of PDE presented in the sequel, namely the
``c system", is not considered at the roots of their models.
Our model, in the end,  comes out pretty much in line with
their approach of general relativity in Cosserat media or
space. Other general relativity models at lower dimensions was
investigated with such approaches as models of gravity   in
\cite{deserjackiwhooft,gottalplert,giddingsabbottkuchar} for
instance.\par\medskip
Also, our present work can be viewed somehow as a kind of generalizing
extension of the T. Fulton
\textit{et al.\/} approach and model \cite{fulton62}.\par\smallskip
{\em Here below, we summarize the mathematical procedure and assumptions 
presented in the sequel and based on the previous discussion (we refer to definitions of 
involution i.e. integrability, symbols of
differential equations, acyclicity and formal integrability such as those
given in  \cite{spencerlin69,gasquigold,pommbook94} for instance):
\begin{enumerate}
\item[$\bullet$] The Riemann scalar curvature $\rho_s$ associated
to the metric field $\underline{\omega}$ hereafter denoted
by $\omega$, is a constant, $n(n-1)k_0$, as a
consequence of the constant Riemann curvature tensor
assumption. In that case, let us recall that the Poincar{\'e}
Lie pseudogroup is involutive (i.e. integrable) since, on the
one hand, its symbol of order 2  vanishes  which makes it
$n$-acyclic, and  on the other hand, it is
formally integrable if $\rho_{\mathrm{s}}$ is a constant. Then,
the Weyl tensor associated to $\underline{\omega}$ is 
vanishing, i.e. we have a conformally flat structure.
\item[$\bullet$] The  system {\eqref{first}} of differential equations
in $\hf_0$, hereafter denoted by $\hf$, being  non-integrable (i.e.
it exists non-analytic
$C^\infty$ solutions on some open neighborhoods), will be supplied
with an other  system of equations (the system we call ``c
system" in the sequel)), obtained from a prolongation procedure
which will be stopped as the integrability conditions of the 
resulting complete system of partial differential equations is
met (i.e. all
$C^\infty$ solutions are analytic almost everywhere).
\item[$\bullet$] The covariant derivations involved in the prolongation
procedure will be assumed to be torsion free, i.e. we will make use of
the Levi-Civita covariant derivations.
\item[$\bullet$] We will extract from the latter system of PDE,
a subsystem, which will be called the ``c system", defining completely
the sub-pseudogroup $\SGammaH$ of those applications $\hf$ which are
strictly smooth deformations of  applications $f$. This PDE
subsystem   will be satisfied by a function $\alpha_0$, hereafter
denoted by $\alpha$. {\em This is the core system of our model and to
our knowledge it has never been really studied or at least related to
any unification  model.\/}
\item[$\bullet$] By considering Taylor series  solutions of
the ``c system", and searching for conditions under which an
$\Rset$-valued Taylor serie $s(x,x_0)$\,\, around a point
$x_0\in\Rset^n$ does not depend on $x_0$, we will show how  general
solutions depend on a particular finite set of parametrizing functions.
\item[$\bullet$] We show that this set of parametrizing functions is
associated to a differential sequence, and that they
can be identified with  both electromagnetic and gravitational
gauge potentials. Moreover, they obey  a set of PDE called the
``first group of PDE". {\em  We  thus get a gauge theory, similar to the
electromagnetic theory, though not based on the de Rham sequence,
but rather on a kind of linear Spencer sequence for non-linear PDE\/}
(see the general linear sequences for the linear PDE case
in~\cite{spencerlin69}).
\item[$\bullet$] We deduce the  metric field
$\nu\equiv\omega+\delta\omega$ of the infinitesimal smooth
deformations, depending on the electromagnetic and gravitational
gauge potentials.
\item[$\bullet$] We give general Lagrangian densities associated to
$\nu$, and in particular we show how the occurence of a spin property
makes it possible to define an electromagnetic
Lagrangian density, sheding some new light on the relation between spin
and electromagnetic interactions with charged particules.
\item[$\bullet$] We conclude with various remarks about physical
interpretations of general relativity.
\end{enumerate}}
To finish this chapter, we indicate that the mathematical tools
used for this unification  finds its roots, first in  the
conformal Lie structure that has  been extensively
studied by H. Weyl \cite{weylkonf}, K. Yano \cite{yanobook70},
J.~Gasqui \cite{gasqui} and J. Gasqui
\& H. Goldschmidt \cite{gasquigold} for instance, and in second place, in
the non-linear cohomology of Lie equations studied by B.
Malgrange \cite{malgrangeI,malgrangeII}, A.~Kumpera
\& D.~Spencer \cite{kumperaspencer} and J.-F.~Pommaret \cite{pommbook94}.
Meanwhile we only partially refer to some of these aspects since it
mainly has to do with the general theory of Lie equations, and not
exactly with the set of PDE we are concerned with. Indeed these set of
PDE is not the set of conformal Lie equations themselves, but rather a
kind of ``residue" coming from a comparison with  ``Poincar\'e Lie
equations", i.e. the ``c system". Also, a large amount of mathematical
results have been obtained and complete reviews exist, that are devoted
to conformal geometries \cite{akivisgoldbergbook}. Hence, most of the results concerning this
geometry are summarized in the sequel. We essentially indicate succintly
the cornerstones which are absolutely necessary for our explanations and
descriptions of the model.


\section{The conformal finite Lie equations of the substratum spacetime}
First of all, and from the previous sections, we assume that
the group of relativity is no longer  Poincar\'{e}  but
the conformal Lie group. In particular, this involves that no
physical law changes occur shifting from a given frame embeded in a
gravitational field to a uniformly accelerated relative isolated one, as
we will recall \cite{page}. This is just the elevator metaphor at the
origin of the Einstein's principle of equivalence.\par The conformal
finite Lie equations are deduced from the conformal action on a {\it
local\/} me\-tric field
$\omega$ defining a  pseudo-Riemannian structure on
$\Rset^n\simeq\Sub$. We insist on the fact we do local studies, 
meaning we consider {\it local\/}
charts from open subsets of the latter manifolds into a
common open subset of $\Rset^n$. Hence by geometric objects
or computations on $\Rset^n$, we mean {\it local\/} geometric
objects or computations  on the
manifolds $\m$, $T\m$ and $\Sub$. Also, it is well-known that the
mathematical results displayed below are independent of the dimension
when greater or equal to 4. Let us consider
${\hf}\in\diff{\infty}(\Rset^n)$, the set of local
diffeomorphisms of $\Rset^n$ of class $C^\infty$, and any
function $\alpha\in{C^\infty}(\Rset^n,{\Rset})$. Then if ${\hf}\in{\GammaG}$
($\GammaG$ being the pseudogroup of local conformal bidifferential
maps on $\Rset^n$), ${\hf}$ is a solution  of the PDE system (in fact
other PDE must be satisfied to completely define $\GammaG$ as one will
be seen in the sequel):
\begin{equation}
\hf^\ast(\omega)=e^{2\alpha}\,\omega\,,
\label{1}
\end{equation}
with $\det(J({\hf}))\not=0$, where
$J({\hf})$ is the Jacobian of $\hf$, and ${\hf}^{\ast}$ is
the pull-back of $\hf$. Also, only the $\e^{2\alpha}$
positive functions are  retained in view of the previous assumption,
that only one orientation is chosen on $\Rset^n$. We therefore 
consider only the $\hf$'s which preserve that orientation.  We  
recall that $\alpha$ {\em is a varying function depending on each $\hf$
and consequently not fixed\/}. We denote
$\tilde{\omega}$ the metric on $\Rset^n$ such as by definition:
$\tilde{\omega}\simeq\e^{2\alpha}\omega$, and we agree on putting a
tilde on each tensor or geometrical ``object" relative to, or
deduced from this  metric $\tilde{\omega}$. 
\par Now, performing a first prolongation of the system \eqref{1}, we
deduce other second order PDE's connecting the Levi-Civita covariant
derivations  $\nabla$ and
$\widetilde{\nabla}$, respectively associated to $\omega$
and $\tilde{\omega}$. These new  differential equations are 
(see for instance \cite{gasqui}) $\forall\, X,\,Y\in
T\Rset^n$:
\begin{equation}
{{\widetilde{\nabla}}_X}Y={\nabla_X}Y+d\alpha(X)Y+d\alpha(Y)X-
\omega(X,Y)\,{{}_\ast}d\alpha\,, \label{3} \end{equation}
where $d$ is the exterior differential and ${{}_\ast}d\alpha$
is the dual vector field of the 1-form $d\alpha$ with respect
to the metric $\omega$, i.e. such that $\forall\,
X\in T\;\Rset^n$:
\begin{equation}
\omega(X,{{}_\ast}d\alpha)=d\alpha(X)\,.
\label{4} \end{equation}
Since the Weyl tensor $\tau$ associated to $\omega$ is
assumed to vanish, the Riemann tensor $\rho$ can be
rewritten
$\forall\,X,Y,Z,U\in{C^\infty}(T\Rset^n)$ as:
\begin{multline} 
\omega(U,\rho(X,Y)Z)={\frac{1}{(n-2)}}
\left\{\omega(X,U)\sigma(Y,Z)-
\omega(Y,U)\sigma(X,Z)\right.\\
\left.+\,\omega(Y,Z)\sigma(X,U)-
\omega(X,Z)\sigma(Y,U)\right\}, 
\label{6} 
\end{multline}
where $\sigma$ is  defined  by (see the tensor ``\,L\," in \cite{yanobook70} up
to a constant depending on $n$)
\begin{equation}
\sigma(X,Y)={\rho_{\mathrm{ic}}}(X,Y)-
{\frac{\rho_{\mathrm{s}}}{2(n-1)}}\omega(X,Y),
\label{7} 
\end{equation}
where $\rho_{\mathrm{ic}}$ is the Ricci tensor and $\rho_{\mathrm{s}}$
is the Riemann scalar curvature. Consequently, the  first
order system of PDE in $\hat{f}$ ``connecting" $\tilde{\rho}$
and $\rho$, can be rewritten as a first order system of PDE
concerning $\tilde{\sigma}$ and
$\sigma$. Using the torsion
free property of the Levi-Civita covariant derivations, one obtains the
\textit{third\/} order system of PDE (since $\alpha$ is depending on the first
order derivatives of $\hf$):
\begin{multline} 
\hf^\ast(\sigma)(X,Y)\equiv\tilde{\sigma}(X,Y)=\sigma(X,Y)+(n-2)\Big(d\alpha(X)
d\alpha(Y)\\
-{\frac{1}{2}}\omega(X,Y)
d\alpha({{}_\ast}
d\alpha)
-\mu(X,Y)\Big),
\label{10} 
\end{multline}
in which we have defined the symetric
tensor $\mu\in{C^\infty}({S^2}\Rset^n)$ by:
\begin{equation}
\mu(X,Y)={\frac{1}{2}}\left[X(\,d\alpha(Y))+Y(d\alpha(X))-
d\alpha({\nabla_X}Y+{\nabla_Y}X)\right]. \label{9}
\end{equation}
To go further, it is important  again to notice that the relation
\eqref{10} is directly related to a third order system of PDE we
denote (T), since it is deduced from a supplementary prolongation
procedure applied to the second order system 
\eqref{3}. Then it follows, from  the well-known theorem of H. Weyl on
equivalence of conformal structures
\cite{weylkonf,yanobook70,gasquigold}, and because of the Weyl tensor
vanishing, that the systems of differential equations
\eqref{1} and \eqref{3}  when completed with the latter
third order system (T), becomes an involutive system of
order three. Let us stress again
that $\alpha$ is  merely  defined by $\hf$ and its first order
derivatives, according to the relation \eqref{1}.\par Looking
only at those applications $\hf$  which are smooth deformations of
applications $f$,  this third order system of PDE must reduce to  a
particular  conformal Lie sub-pseudogroup we denote by $\SGammaH$.
Indeed, if $\alpha$ tends towards the zero function with respect to the
$C^2$-topology, then the previous set of smoothly deformed applications
$\hf$ must tend towards the Poincar{\'e} Lie pseudogroup. But this
condition is not satisfied by all of the application $\hf$ in the
conformal Lie pseudogroup $\GammaG$, since in full generality, the
non-trivial third order system of PDE (T) would be kept at the zero
$\alpha$ function limit. In other words, the pseudogroup $\SGammaH$
could be defined by an involutive second order system of PDE which would
tend towards the involutive system defining the Poincar{\'e} Lie
pseudogroup.\par The systems of differential equations \eqref{1} and
\eqref{3} would be well suited to define partially this pseudogroup
$\SGammaH$. The
$n$-acyclicity property of $\SGammaH$ would be restored and borrowed,
at the order two, from the  Poincar{\'e} one, provided however that  an,
a priori, arbitrary  \emph{input\/} perturbative function $\alpha$ is
given before, instead of being defined from an application $\hf$
according to the relation \eqref{1}. But the formal integrability is
obtained  only if  the tensor $\sigma$  satisfies one of the following 
equivalent relations (see formula (16.3) with definition (3.12) in 
\cite{gasquigold}):
\begin{equation} 
\sigma={k_0}{\frac{(n-2)}{2}}\,\omega\quad
\Longleftrightarrow\quad\rho_{\mathrm{ic}}=(n-1)\,k_0\,\omega\,,
\label{11}
\end{equation}
deduced from relation \eqref{7}, in order to avoid adding up
the supplementary  first order system of PDE \eqref{10} to \eqref{1}.
Then, considering the system \eqref{1}, the system (\ref{10}) reduces
to  a second order system of PDE  concerning only the input function
$\alpha$, which is thus constrained, contrarily to what might have been
expected, and such that:
\begin{equation} 
\mu(X,Y)={\frac{1}{2}}\left\{
\left[{k_0}\left(1-{\e^{2\alpha}}\right)
-d\alpha({{}_\ast}d\alpha)
\right]\omega(X,Y)\right\}
+d\alpha(X)d\alpha(Y)\,. 
\label{12} 
\end{equation}
Obviously, as can be easily verified, this is an involutive system of PDE
as it can be easily verified, since it  is a formally integrable system
with a vanishing symbol (i.e. elliptic symbol) of order two.\par Thus,
we have  series of PDE deduced from \eqref{1} defining all the
\emph{smooth\/} deformations of the applications $f$ contained in the
conformal Lie pseudogroup $\GammaG$.\par In addition, the metric field
$\omega$ necessarily satisfies relation
\eqref{11} (which is analogous to the Einstein equations but with a
stress-energy tensor proportional to the metric one) if  the Lie
pseudogroup
$\SGammaH$, containing the Poincar{\'e} one, strictly differs from the
latter. In that case, such a metric field
$\omega$ and the substratum spacetime $\Sub$ will be called
\mbox{\textit{S-admissible\/}} (with S as ``Substratum"), and this
S-admissibility is assumed to be satisfied  in the
sequel.\par   
In an orthonormal system of coordinates, the PDE
\eqref{1}, \eqref{3} and \eqref{12} defining
$\SGammaH\subset\GammaG$ can be written, with
$\det(J({\hf}))\not= 0$ and $i,j,k=1,\cdots,n$ as:
\begin{subequations}
\begin{align}
&\sum_{r,s=1}^n\omega_{rs}(\hf)\,\,\hf^{r}_{i}\hf^{s}_{j}=
e^{2\alpha}\omega_{ij}\,,
\label{f1}\\ 
&\hf^k_{ij}+
\sum_{r,s=1}^n{}\gamma^{k}_{rs}(
\hf)\,\,\hf^{r}_{i}\hf^{s}_{j}=
\sum_{q=1}^n{}\hf_q^k\left(
\gamma^{q}_{ij}+\alpha_i\delta^q_j+
\alpha_j\delta^q_i-\omega_{ij}\alpha^q\right)\,,
\label{f2}\\
&\mu_{ij}=\alpha_{ij}-
\sum_{k=1}^n\alpha_k\gamma^k_{ij}=\frac{1}{2}
\Big\{k_0(1-\mbox{e}^{2\alpha})-
\sum_{k=1}^n\alpha^k\alpha_k
\,\Big\}\omega_{ij}+
\alpha_i\alpha_j\,,\label{third}
\end{align}
\label{confeq}
\end{subequations}
where $\delta^i_j$ is the Kronecker tensor, and where one
denotes as usual
 $\hf^i_j\equiv\partial \hf^i/\partial
x^j\equiv\partial_j \hf^i$, etc~\dots, $T_k=\sum_{h=1}^n
T^h\,\omega_{hk}$ and $T^k=\sum_{h=1}^n
T_h\,\omega^{hk}$ for any tensor $T$ where $\omega^{ij}$
is the inverse metric tensor, and
$\gamma$ is the  Riemann-Christoffel form associated to the
\emph{S-admissible\/} metric $\omega$. This is the set of our
starting equations. It matters to notice that the (T) system is not
included in the above set of PDE. Indeed, the latter being
already involutive from order 2, this involves from definition of
involution that the (T) system is redundant since all the applications
$\hf\in\SGammaH$, solutions of \eqref{confeq}, will also be solutions of
all the  systems of PDE obtained by prolongation.
\par Under a change of coordinates with a conformal application $\hf$ and
assuming a constant metric  field $\omega$, the function
$\alpha(\equiv\tilde{\alpha}_0)$ and the tensor
$\tilde{\alpha}_1\equiv\{\alpha_1,\cdots,\alpha_n\}$ are
transformed into ``primed" functions and tensors such as
($j=1,\cdots,n$):
\[
\alpha'(\hf)=\alpha-\frac{1}{n}\ln{|}\det
J(\hf)\,{|}\,,\qquad
\sum_{i=1}^{n}\hf^i_j\,{\alpha}_i'(\hf)=\alpha_j-\sum_{k,l=1}^{n}
\frac{1}{n}\,(\hf^{-1})^{k}_l\circ\hf\,\,\hf^l_{kj}\,,
\]
which  essentially displays the affine structure of these
``geometrical objects". In particular,  the tensor
$\tilde{\alpha}_1$ can be associated with the second order  derivations
of $\hat{f}$. And Thus, it could be considered as an acceleration
tensor.\par More precisely, let  $x\in\Rset^n$ be the value of a
differential application $\phi(\tau)$ depending on a real parameter
$\tau$ such that 
$\phi(0)\equiv x_0$, and assuming  $\hf(x_0)=x_0(\equiv 0$ from
\eqref{tilom}). Then, denoting by a
dot ``$\,\,\,\dot{\,}\,\,\,$" the derivative with respect to $\tau$,
assuming $\hx$ and $x$  are such that $\hx=\hf(x)$ and 
${\parallel}\dot{x}{\parallel}^2\equiv\omega(\dot{x}\,,\dot{x})=1$,
 and considering a composed application
$\hat{g}\,\circ\hf\in\GammaG$ instead of $\hf$ alone, we
easily deduce from \eqref{f1} the relation:
\[
\alpha'(\hx)=\alpha(x)-\ln
({\parallel}\dot{\hx}{\parallel})\,.
\]
And by differentiating the latter, we obtain also:
\[
{\alpha}_i'(\hx)=\sum_{j=1}^{n}
(\hf^{-1})^j_i\circ\!\hf(x)\,\,\alpha_j(x)-
\frac{\ddot{\hx}_i}{{\parallel}\dot{\hx}{\parallel}^2}\,.
\]
In particular, if $\tau=0$ and ${\parallel}\dot{\hx}{\parallel}^2=1$,
then the Jacobian matrice of $\hf$ at $x_0$ is a Lorentzian matrice and
\begin{equation}
\alpha'(x_0)=\alpha(x_0)\,,\qquad{\alpha}_i'(x_0)=\sum_{j=1}^{n}
(\hf^{-1})^j_i(x_0)\,\,\alpha_j(x_0)-
\ddot{\hx}_i\,.
\label{axel}
\end{equation}
But from relations \eqref{tilom}, $\hf$ can also be associated to
composed diffeomorphisms such as for instance
$\varphi_{p_0}^{-1}\circ{\varphi}_{p_0}'$, where ${\varphi}_{p_0}'$ 
and ${\varphi}_{p_0}$ define two different local equivalences at $p_0$. 
Then, the points $x$ and $\hx$ can be ascribed, respectively, 
to moving points $p$ and $\hat{p}$ in the vicinity of $p_0$. 
In that case, from the latter geometrical and/or physical 
interpretation and relations \eqref{axel}, 
a change of acceleration would keep 
physical laws invariant since it would be only associated to a change 
of coordinates. It would be close to Einsteinian relativity and to the
Einstein's principle of equivalence. Indeed the ``elevator metaphor" in this
principle, points out the affine feature of the acceleration which
involves equivalence between relative uniformly accelerated
frames with relative velocity $\dot{\hx}$ and relative acceleration
$\ddot{\hx}$ at their crossing point $x_0$. This extends the results
obtained first by J. Haantjes \cite{haantjes40}.\par Consequently,
on the one hand, it is important to notice that $\mu$ or equivalently the
tensor $\tilde{\alpha}_2\equiv\{\alpha_{ij},i,j=1,\cdots,n\}$ might be
considered as an Abraham-E\"{o}tv\"os type tensor
\cite{abraham,misner} encountered in the  E\"otv\"os-Dicke
experiments for the measurement of the stress-energy tensor of the
gravitational potential.\par At this point,  a first set of 
physical interpretations (up to a constant for units and with $n=4$) may
be devised: The tensor
$\mu$ may be ascribed to the stress-energy tensor, $\tilde{\alpha}_1$ to
the gravity acceleration 4-vector  and
$\alpha$ to the Newtonian potential of gravitation.\par On the other
hand, following W. M. Tulczyjew \cite{tulcz83} and J.-F. Pommaret \cite{pommbook94}, $\alpha$ being 
associated with  dilatations it might be
considered as a relative Thomson type temperature:
\(
\alpha=\ln({T_0}/T)\,, 
\)
where $T_0$ is a constant temperature of reference related
with the substratum spacetime $\Sub$. A first question
arises about this temperature $T_0$: Can we consider
it as the 2.7~K cosmic relativistic invariant background temperature ? 
Also, in view of the transformation law of this function, this would
imply that the temperature $T$ is not  a conformal invariant but
only a Lorentz  invariant (as the 2.7 K relic radiation)  since in that
case $|\det J(\hat{f})|=1$ and $\ddot{\hx}=0$ (see also the
Helmotz Lorentz  invariant representation of the absolute temperature
\cite{debrogliebook64} related to a A. H. Taub variational
principle \cite{tulcz83}, and the two opposite non-Lorentz
invariant temperature transformation laws given on the one hand by A.
Einstein \cite{einsteintemp1907}, M. von Laue \cite{vonlauetemp1907},  M.
Planck \cite{plancktemp1908}, and on the other hand by H. Z. Ott
\cite{otttemp1963}, H. Arzeli{\`e}s \cite{arzeliestemp1965}, C.
M{\o}ller \cite{mollertempbook71}).\par In fact, 
$\alpha$ could be considered  either as a Newtonian 
potential of gravitation or as a  temperature  (if 
physically valid),   or may be more judiciously as a sum of
the two. It appears that physical interpretations could be
made at two scales: a ``global" one at universe scale ($\Sub$),  
and a ``local" one ($\tom$) in considering local forces of gravitation
attached to a point $p_0\in\m$. In the same way, a second question
arises at the ``global" scale about the meaning of the tensor
$\tilde{\alpha}_1$  up to units: Since in this case it might be an
acceleration (or a gradient of entropy) of reference associated to
$\Sub$, could it be the acceleration of an inflation process (or a time
arrow) ?  As we shall see, the
various dynamics will  depend only on the physical interpretations of
these two sets of parameters
$\alpha$ and
$\tilde{\alpha}_1$.


\section{The functional dependency of the spacetime deployment}
Now we look  for the formal series, solutions of the system of
PDE (\ref{confeq}), assuming from now that the metric $\omega$ is 
analytic. We know these series will be convergent in a suitable open
subset and thus providing analytic solutions,  since the analytic system
is involutive and in particularly elliptic because of a vanishing symbol
(see, in appendix 4 of \cite{malgrangeII}, the Malgrange theorem for
elliptic systems, given without the assumption of existence of flags of 
Cauchy data, and generalizing the Cartan-K\"ahler theorem
\cite{cartankal,bryantchern} on analyticity; see also the chapters about
the $\delta$-estimate tool  for elliptic symbols in
\cite{spencerlin69}; \cite{rustreid}). Nevertheless we need of course
to know the Taylor coefficients. For instance we can choose  for the
applications
$\hat{f}$ and the functions
$\alpha$  the following series at a point
$x_0\in\Rset^n$:
\begin{align*}
&\hf^i(x):\,
S^i(x,x_0,\{\hat a\})=\sum_{|J|\geq0}^{+\infty}
\hat{a}^i_J(x-x_0)^J/|J|!\,,\\
&\alpha(x):\,
s(x,x_0,\{c\})=\sum_{|K|\geq0}^{+\infty} 
c_K(x-x_0)^K/|K|!\,,
\end{align*}
with $x\in U({x_0})\subset\Rset^n$ being a suitable open 
neighborhood of $x_0$ to insure the convergence of
the series,
$i=1,\cdots,n$, $J$ and
$K$ are multiple index notations such as $J=(j_1,\cdots,j_n)$, 
$K=(k_1,\cdots,k_n)$ with $|J|=\sum_{i=1}^n j_i$ and
similar expressions for $|K|$. Also $\{\hat{a}\}$ and
$\{c\}$ are the sets of Taylor coefficients and
$\hat{a}^i_J$ and $c_K$ are real values and not functions
of $x_0$, though of course, they can also be values of functions at
$x_0$.


\subsection{The ``c system"}
We call the ``c system", the system of PDE
(\ref{third}) (see \cite{haantjes41} for an analogous system with $k_0=0$). 
It is from this set of PDE that  gauge potentials
and fields of interactions could occur. From the series $s$,
at zero-th order one obtains the algebraic equations
($i,j=1,\cdots,n$; $\bfc_1=\{c_1,\cdots,c_n\}$):
\begin{equation}
c_{ij}=\frac{1}{2}\Big\{k_0(1-\e^{2c_0})-
\sum_{k,h=1}^n\omega^{kh}(x_0)c_h c_k
\,\Big\}\omega_{ij}(x_0)+
c_i c_j+\sum_{k=1}^n\,c_k\gamma^k_{ij}
\equiv
F_{ij}(x_0,c_0,\bfc_1)\,,
\label{c2eq}
\end{equation}
and it follows that the $c_K$'s such that $|K|\geq2$, will
depend recursively only on $x_0$, $c_0$ and $\bfc_1$.
It is none but the least the meaning of formal integrability of
so-called involutive systems. Hence the
series for $\alpha$ can be written as convergent series
$s(x,x_0,c_0,\bfc_1)$ with respect to powers of $(x-x_0)$, $c_0$ and
$\bfc_1$. Let us notice that we can change or not the
function~$\alpha$ by varying $x_0$, $c_0$ or
$\bfc_1$.\par
Let $J_1$ be the 1-jets affine bundle of the
$C^\infty$ real valued functions on $\Rset^n$. Then it exists
a subset associated to
$\bfc_0^1\equiv(x_0,c_0,\bfc_1)\in J_1$, we denote by
$\Su(\bfc_0^1)\subset J_1$, the set of elements
$({x}_0',{c}_0',\bfc_1')\in J_1$, such that there
is an open neighborhood
$U(\bfc_0^1)\subset \Su(\bfc_0^1)$, projecting on $\Rset^n$ in an open
neighborhood of a given
$x\in U(x_0)$, for which for all $({x}_0',{c}_0',\bfc_1')\in
U(\bfc_0^1)$ then
$s(x,x_0,c_0,\bfc_1)=s(x,{x}_0',{c}_0',\bfc_1')$.
Assuming the variation $ds$ with respect
to $x_0$, $c_0$ and $\bfc_1$ is vanishing, at a given fixed $x$, is the
subset $\Su(\bfc_0^1)$ a submanifold of $J_1$ ? From $ds\equiv0$ it
follows that ($k=1,\cdots,n$):
\begin{eqnarray*}
\sg_0\equiv dc_0-\sum_{i=1}^nc_i\,dx_0^i=0\,,\qquad\sg_k\equiv dc_k-
\sum_{j=1}^n F_{kj}(x_0,c_0,\bfc_1)\,dx_0^j=0\,.
\end{eqnarray*}
We recognize a regular analytic Pfaff system  we denote
$P_c$, generated by the 1-forms $\sg_0$ and $\sg_k$, and
the meaning of their vanishing is that the solutions
$\alpha$ do not change for such variations of $c_0$,
$\bfc_1$ and $x_0$. Also, as can be easily
verified, the Pfaff system $P_c$ is integrable since the
Fr\"obenius condition of involution is satisfied, and all the
prolongated 1-forms $\sg_K$ ($|K|\geq2$) will be linear
combinations of these $n+1$ generating forms thanks to the
recursion property of formal integrability. Then the subset
$\Su(\bfc_0^1)$ of dimension $n$ containing a particular
element $\bfc_0^1\equiv(x_0,c_0,\bfc_1)$ is a submanifold of
$J_1$. It is a particular leaf of, at least, a local foliation on $J_1$
of codimension $n+1$.\par Since the system of PDE defined by the
involutive Pfaff system $P_c$, namely the c system, is elliptic (i.e.
vanishing symbol) and formally integrable, one deduces that it exists on
$J_1$, local systems of coordinates $(x_0,\tau_0,\tau_1,\cdots,\tau_n)$
such that each leaf $\Su(\bfc_0^1)$ is an analytic submanifold
\cite{malgrangeII} for which $\tau_0=cst$ and $\tau_i=cst$ $(i=1,\cdots
,n)$. This involves that all the convergent series
$s(x,{x}_0',{c}_0',\bfc_1')$ with
$({x}_0',{c}_0',\bfc_1')\in\Su(\bfc_0^1)$ equal one only analytic
solution function $u(x,\tau_0,\pmb{\tau}_1)$ 
($\pmb{\tau}_1\equiv\{\tau_1,\cdots,\tau_n\}$), analytic  with
respect to $x$  as well as  with respect to the
$\tau$'s. This results of the continuous series $s$ which are convergent
whatever the fixed set of given values $x$, $x_0$, $c_0$  and
$\bfc_1$. Thus, in full generality, considering the difference
$s(x,x_0,c_0,\bfc_1)-s(x,{x}_0',{c}_0',\bfc_1')$
we have the relation:
\begin{equation}
s(x,x_0,c_0,\bfc_1)-s(x,{x}_0',{c}_0',\bfc_1')=
u(x,\tau_0,\pmb{\tau}_1)-
u(x,{\tau}_0',\pmb{\tau}_1')\,,
\label{diffs}
\end{equation}
with  $\tau$ parameters related by ($i=1,\cdots,n$)
\begin{equation}
\triangle_0\tau\equiv{\tau}_0'-\tau_0=
\int_{\bfc^1_0}^{{\bfc'}^1_0}\!
\sg_0\,,\qquad\triangle_i\tau\equiv{\tau}_i'-\tau_i=
\int_{\bfc^1_0}^{{\bfc'}^1_0}\!
\sg_i\,.
\label{integsig}
\end{equation}
Now, we consider the
$c$'s are values of differential (i.e. $C^\infty$) functions
$\rho\,$: $c_K=\rho_K(x_0)$, as expected for Taylor series
coefficients, and defined on a starlike open neighborhood of $x_0$.
Roughly speaking, we make a pull-back on $\Rset^n$ by differentiable
sections, inducing a projection from the subbundle of projectable
elements in
$T^\ast\!J_1$ to
$T^\ast\Rset^n\otimes_\Rset~\!J_1$. Then, we set 
(with $\pmb{\rho}_1\equiv\{\rho_1,\cdots,\rho_n\}$
and no changes of notations for the pull-backs):
\begin{subequations}
\begin{align}
&\sg_0\equiv\sum_{i=1}^n(\partial_i\rho_0-\rho_i)\,dx_0^i\equiv
\sum_{i=1}^n\A_i\,dx_0^i\,,
\label{rhoA}\\
&\sg_i\equiv\sum_{j=1}^n(\partial_j\rho_i-
F_{ij}(x_0,\rho_0,\pmb{\rho}_1))\,dx_0^j
\equiv\sum_{j=1}^n\B_{j,i}\,dx_0^j\,,
\end{align}
\label{rhoAB}
\end{subequations}
and it follows the integrals \eqref{integsig} must be
performed from $x_0$ to ${x}_0'$ in a starlike open neighborhood of
$x_0$. In particular, if
$\bfc_0^1$ is an element of  the ``null" submanifold corresponding to
the vanishing solution of the ``c system", then 
the difference (\ref{diffs}) involves that
\begin{equation*}
\alpha(x)\equiv
s(x,{x}_0',{c}_0',\bfc_1')=
u(x,{\tau}_0',\pmb{\tau}_1')\,,
\end{equation*}
with
\[
{\tau}_0'=
\int_{x_0}^{{x}_0'}
\sum_{i=1}^n\A_i\,dx^i+\tau_0\,,\qquad{\tau}_i'=
\int_{x_0}^{{x}_0'}
\sum_{j=1}^n\B_{j,i}\,dx^j+\tau_i\,.
\]
Then we deduce:
\begin{theorem}
All the analytic solutions of the involutive system of PDE\,
{\eqref{third}} can be written in a suitable starlike open
neigborhood of $x_0$ as
\begin{equation}
\alpha(x)\equiv u\Big(x,\int_{x_0}^{x'_0}
\sum_{i=1}^n\A_i\,{dx'}^i+\tau_0\,,\int_{x_0}^{x'_0}
\sum_{j=1}^n\B_{j,1}\,{dx'}^j+\tau_1
,\cdots,\int_{x_0}^{x'_0}
\sum_{j=1}^n\B_{j,n}\,{dx'}^j+\tau_n\,\Big)\,,
\label{deformpar}
\end{equation}
with
\(u(x_0,\tau_0,\tau_1,\cdots,\tau_n)=0\), $x'_0\in U(x_0)$ and
where $u$ is a unique fixed analytic function depending on the
$\mbox{n(n+1)}$ $C^\infty$ functions
$\A_i$ and $\B_{j,k}$ defined by the relations {\eqref{rhoAB}}.
The integrals in $u$ are called the ``potential of
interactions". Let us remark that we can set $x'_0\equiv
v(x)$ if the gradient of $v$, i.e. $\nabla v$, is
in the annihilator of the Pfaff system $P_c$ of
1-forms $\sigma$.\end{theorem} Also this result shows the
functional dependencies of the solutions of the ``c system"
with respect to the functions
$\rho_0$ and
$\pmb{\rho}_1$, themselves associated to the smooth
infinitesimal deformations of these solutions. These  smooth
infinitesimal deformations gauge fields $\A$ and $\B$, defined
by $n(n+1)$ potential functions ($\bf 20$ functions if $n=4$),
can also be considered as infinitesimal smooth deformations
from ``Poincar{\'e} solutions" of the system (\ref{confeq}) for
which
$\alpha\equiv0$, to some ``conformal solutions" whatever is
$\alpha$.\par Moreover the functions
$\rho$, and consequently the functions
$\rho_0$, $\A$ and
$\B$, must satisfy additional differential equations
coming from  Fr\"obenius conditions of involution of the Pfaff
system $P_c$. More precisely, from the relations
$d\sg_0=\sum_{i=1}^n\,dx_0^i\wedge\sg_i$,
$d\sg_i=\sum_{j=1}^n\,dx_0^j\wedge\sg_{ij}$ and 
\begin{equation}
\sg_{ij}=c_i\sg_j+c_j\sg_i-\omega_{ij}\Big\{k_0\e^{2c_0}\sg_0+
\sum_{k,h=1}^n\omega^{kh}c_h\sg_k\Big\}
+\sum_{k=1}^{n}\gamma^k_{ij}\sigma_k
\equiv\vartheta_{ij}(\bfc_0^1,\sg_J;|J|\leq1)\,,
\label{sg2}
\end{equation}
one deduces
a set of algebraic relations at $x_0$:
\begin{subequations}
\begin{equation}
\J_{k,j,i}\equiv
\omega_{ij}\left\{k_0e^{2\rho_0}\A_k+
\sum_{r,s=1}^n
\omega^{rs}\,\rho_r\,\B_{k,s}\right\}
+\partial_j\B_{k,i}-
\rho_i\,\B_{k,j}-
\rho_j\,\B_{k,i}
-\sum_{s=1}^{n}\gamma^s_{ij}\B_{k,s}\,,
\label{JAB}
\end{equation}
\begin{equation}
\I_{i,k}=\I_{k,i}\equiv\partial_k\A_i-\B_{i,k}\,,
\qquad \J_{k,j,i}=\J_{j,k,i}\,.\label{IAB}
\end{equation}
\label{ABIJeq}
\end{subequations}  
Clearly, in these relations, the set of functions
$(\rho_0,\pmb{\rho}_1)$ appears to be a set of arbitrary
differential functions. In considering 
$\F$ and $\G$ as being respectively the skew-symmetric and the
symmetric parts of the tensor of components
$\partial_i\rho_j$, then one deduces, from the symmetry
properties of the latter relations, what we call {\it the
first set of differential equations}:
\begin{subequations}
\begin{align}
&\partial_i\F_{jk}+\partial_j\F_{ki}+\partial_k\F_{ij}=0\,,
\label{max1}\\
&2\,\partial_j\G_{ki}-\partial_i\G_{kj}-\partial_k\G_{ij}=
\partial_i\F_{jk}-\partial_k\F_{ij}\,,
\end{align}
\end{subequations}
with  
\begin{subequations}
\begin{align}
&\F_{ij}=\partial_j\rho_i-\partial_i\rho_j
=\partial_i\A_j-\partial_j\A_i\,,
\label{set1}\\
&\G_{ij}=-(\partial_i\rho_j+\partial_j\rho_i)\equiv
\partial_i\A_j+\partial_j\A_i
\mod(\rho_0,\pmb{\rho}_1)\,.
\end{align}
\label{FG}
\end{subequations}
The PDE \eqref{max1} with \eqref{set1}  might be interpreted as the
first set of {\it Maxwell equations}. In view of physical
interpretations, we can easily  compute the Euler-Lagrange equations
of a conformally equivariant Lagrangian density
\begin{equation}
\Lag(x_0,\rho_0,\pmb{\rho}_1,\A,\F,\G)\,d^n\!x_0\,,
\label{LAB}
\end{equation}
with $\A$, $\F$ and $\G$ satisfying the relations
\eqref{rhoAB} and \eqref{FG}. We would obtain easily what we call {\it
the second set of differential equations}.\par Then we give a
few definitions to proceed further.
\begin{definition} We denote:\par
\begin{enumerate}
\item $\theta_\Rset$, the presheaf of rings of germs of
the differential (i.e.
$C^\infty$) functions defined on $\Rset^n$, 
\item $\underline{J_1}$, the presheaf  of
$\theta_\Rset$-modules of germs of differential sections of
$J_1$,
\item $\Sz\subset \theta_\Rset$, the presheaf  of rings
of germs of functions which are solutions with their first derivatives, of the 
``algebraic equations" GHSS (\ref{third}) taken at any given points
$x_0$ in $\Rset^n$, \underline{not simultaneously} at each point in $\Rset^n$ (see
\textit{Remark 1\/} below),
\item $\Su\subset\underline{J_2}$,  projectable on
$\underline{J_1}$  ($\underline{J_1}\simeq\Su$), the embedding
in $\underline{J_2}$ of the presheaf of
$\theta_\Rset$-modules of germs of differential sections of
$J_2$, defined by the  system 
(\ref{third}) of algebraic equations at any given points
$x_0\in\Rset^n$ (not everywhere, as mentioned above),
\item $\tsms$, the presheaf of  
$\theta_\Rset$-modules of germs  of
global \mbox{1-forms} on $\Rset^n$.
\end{enumerate}
\end{definition}
Remark 1: Through this set
of definitions, we do not consider  PDEs solutions, but instead, solutions of
algebraic equations at any given point $x_0$. In this light, PDEs solutions are
to be regarded as particular ``coherent" subsheafs for which equations
(\ref{third}) are satisfied everywhere in $\Rset^n$, i.e., at $x\neq x_0$, and not
solely at  $x_0$. We insist that the algebraic equations
(\ref{third}) do not concern solutions of a PDE system, but the
values of  second derivatives of  functions at $x_0$, depending on
those of first order at most at $x_0$, with no  constraints between
first and zero-th order values of these functions at $x_0$.
\par\medskip
Then,  considering the local diffeomorphisms
\[(\wedge^k\,T^\ast\Rset^n\otimes_\Rset J_r)_{x_0}
\simeq(\{x_0\}\otimes_\Rset
J_r)\times(\wedge^k\,T^\ast_{\!x_0}\Rset^n
\otimes_\Rset J_r)\] 
with $0\leq k\leq n$ and $r\geq0$,
we set the definitions:
\begin{definition}We define the local operators:\par
\begin{enumerate}
\item
$j_1:(x_0,\rho_0)\in\Sz\longrightarrow
(x_0,\rho_0,\boldsymbol\rho_1,\boldsymbol{\rho}_2)\in\Su$
with 
$\boldsymbol\rho_1=
(\partial_1\rho_0,
\cdots,
\partial_n\rho_0)$ and
$\boldsymbol\rho_2=
(\partial_{11}^2\rho_0,
\partial_{12}^2\rho_0,
\cdots,
\partial_{nn}^2\rho_0)$,
\item 
$D_{1,c}:\boldsymbol{\rho}^2_0\equiv(x_0,\rho_0,
\boldsymbol{\rho}_1,\boldsymbol{\rho}_2)\in\Su
\longrightarrow(\boldsymbol{\rho}^1_0,\sg_0,
\sg_1,\cdots,\sg_n)
\in\tsms\otimes_{\theta_\Rset}
\underline{J_1}$,
with $\A$, $\B$ and $\boldsymbol{\rho}^1_0\equiv(x_0,\rho_0,
\boldsymbol{\rho}_1)$ satisfying
relations (\ref{rhoAB}), and $P_c=\{\sg_0,
\sg_1,\cdots,\sg_n\}$ being a Pfaffian system of linearly independent
regular 1-forms on $J_1$,
\item 
\(D_{2,c}:(\boldsymbol{\rho}^1_0,\sg_0,\sg_1,\cdots,\sg_n)
\in\tsms\otimes_{\theta_\Rset}
\underline{J_1}\longrightarrow
(\boldsymbol{\rho}^1_0,\zeta_0,\zeta_1,\cdots,\zeta_n)\in\)
\(\wedge^2\tsms\otimes_{\theta_\Rset}\underline{J_1}\), with
\begin{eqnarray*}
\zeta_0=\sum_{i,j=1}^n\I_{i,j}\,dx_0^i\wedge\,dx_0^j\,,\qquad
\zeta_k=\sum_{i,j=1}^n\J_{j,i,k}\,dx_0^i\wedge\,dx_0^j\,,
\end{eqnarray*}
the functions
$(x_0,\rho_0,\boldsymbol{\rho}_1,\boldsymbol{\rho}_2)\in\Su$
and the tensors
$\I$, $\J$, $\A$ and $\B$  satisfying the relations (\ref{ABIJeq}).
\end{enumerate}
\end{definition}
Then from all that preceeds, we can deduce:
\begin{theorem}
The differential sequence
\[
\begin{CD}
0@>>>\Sz@>j_2>>\Su
@>D_{1,c}>>\tsms\otimes_{\theta_\Rset}
\underline{J_1}@>D_{2,c}>>
\wedge^2\tsms\otimes_{\theta_\Rset}\underline{J_1}\,,
\end{CD}
\]
with the $\Rset$-linear local
differential operators $D_{1,c}$ and $D_{2,c}$\,,
is  exact (where the first injectivity, namely $j_1$, results from remark 1).
\end{theorem}
Remark 2: Before proceeding with the proof of this Theorem, a
few comments are in order.
{\rm
The continuation ``on the right" of the differential sequence  above
would require, in order to demonstrate the exactness, a
generalization of the Fr\"obenius Therorem to
$p$-forms with $p\geq2$, which, to our knowledge at least, is not
available in full generality, no more as the concept of canonical
contact $p$-forms representations. Indeed, the higher local
differential operators $D_{i\ge3,c}$ would be non-linear, in
contrary to the usual Spencer differential operators, because
of the non-linearity of the GHSS system. We are faced to the
same situation encountered in the Spencer sequences for Lie
equations, these sequences being truncated at this same order
two.} {\rm The sequence above is a physical gauge sequence, for
which we can make the following identification:
$\tsms\otimes_{\theta_\Rset}
\underline{J_1}$ is the space of the gauge \textit{potentials\/} $\A$ and $\B$, whereas
$\wedge^2\tsms\otimes_{\theta_\Rset}\underline{J_1}$ is the space of the gauge \textit{
strength fields\/}
$\I$ and $\J$.}
\par\medskip
This sequence is close to a kind of Spencer
linear sequence \cite{spencerlin69}. It differs essentially
in the tensorial product which is taken on  $\theta_\Rset$
(because of the non-linearity of the ``GHSS system", inducing
a $\rho$ ``dependence" of the various Pfaff forms) rather
than on the $\Rset$ field as is in the original linear Spencer
theory \cite{spencerlin69} (other developements have
included the $\theta_\Rset$ case after this first
Spencer original version). Also, since the system
$P_c$ is integrable, it is always, at least locally, diffeomorphic  to an
integrable set of Cartan 1-forms in
$T^\ast\Rset^n\!\otimes_\Rset\!{J_1}$
associated to a particular finite Lie algebra $g_c$ 
(of dimension greater or equal to
$n+1$), with corresponding Lie group
$G_c$ acting on the left on each leaf of the foliation $\mathcal{F}_1$
\cite{AT92,tT90}.  It follows  that the integrals in
(\ref{deformpar}) would define  a deformation class in the
first non-linear  Spencer cohomology space of deformations of
global sections from
$\Rset^n$ to a sheaf of Lie groups  $G_c$ \cite{kumperaspencer}
(see also \cite{lislereid}, though within a different approach).\par In
addition, 1) in Theorem 1, the fonction $u$ is defined with integrals
associated to the definition of a homotopy operator of the differential
sequence above \cite{buttin67}, and 2) in  Theorem 2, the metric
$\omega$ is allowed to be of class $C^\infty$, rather than analytic, as in 
Theorem 1, because  formal properties only are
considered.\par\medskip  
\textbf{Proof of Theorem 2:} At $\Sub_c^1$ the sequence
exactness is trivial and we may pass to the exactness of the
differential sequence at
$\tsms\otimes_{\theta_\Rset}\underline{J_1}$.\par
In a neighbourhood of an open set $V(\bfcM)\subset J_1$ of
$\bfcM\in J_1$, the condition $D_{2,c}(\sg)=0$ implies the
relations:
\begin{eqnarray}
&d\sgt_0=\sum_{i=1}^n\,dx_0^i\wedge\sgt_i\,,\label{dsg0}\\
&d\sgt_i=\sum_{j=1}^n\,dx_0^j\wedge\sgt_{ij}\label{dsgi}
\end{eqnarray}
with:
\begin{equation}
\sgt_{ij}=c_i\sgt_j+c_j\sgt_i-\omega_{ij}\Big\{k_0 e^{2c_0}\sgt_0+
\sum_{k,h=1}^n\omega^{kh}c_h\sgt_k\Big\}
+\sum_{k=1}^{n}\gamma^k_{ij}\sgt_k\,,
\label{fhs}
\end{equation}
the ``$\tilde\sg$" 1-forms  are defined above $J_1$, and
correspond to the 1-forms $\sg$ defined in a  neighbourhood
$W(X_0)\subset\Rset^n$ at $x_0\in W(X_0)$: they are such that if
$p_1: J_1\longrightarrow\Rset^n$ stands for the standard projection,
then $p_1(V(\bfcM))=W(X_0)$, $p_1(\bfcm)=x_0$ and
$p_1(\bfcM)=X_0$.\par Regularity and
linear independence, ensure the existence of a locally integrable
manifold $\mathcal{V}_1$, with dimension $n$, and of 
$n+1$ first integrals $\{y_\nu\}$
($\nu=0,1,\dots,n$). Up to  constants, the functions $y_\nu$
can be choosen such that
$y_\nu(\bfcM)=0$. Then, at
$\bfcM$, we have  the relations:
\begin{equation}
\tilde{\sg}_\nu(\bfcM)\equiv dy_\nu\!\big/_{\!\!\bfcM}\,,
\end{equation}
and in $V(\bfcM)$, the relations 
\begin{equation}
\tilde{\sg}_\nu=dy_\nu-\sum_{\mu\neq\nu}^{n} f_\nu^\mu(y)dy_\mu\,,
\label{desy}
\end{equation}
with $f_\nu^\mu(y)\longrightarrow0$ when $y\longrightarrow0$, that
is when $\bfcm\longrightarrow\bfcM$.\par These relations can also
be defined on the presheaves of the $J_1$ local
sections. This is because it exists a $C^1$-mapping, say $s$,
from  $W(X_0)$ into $V(\bfcM)$, such that
$s(W(X_0))=U(\bfcM)\subset V(\bfcM)$ and
$s(x_0)=\bfcm$. And thus locally, one has
$\mathcal{V}_1\cap U(\bfcM)\simeq W(X_0)$. In the relations
(\ref{desy}), it is therefore possible to take
$y_j\equiv c_j$ ($j=1,\dots,n$). Setting
$s^\ast(dy_j)=s^\ast(dc_j)\equiv dx^j_0$, and denoting by ``$\rho$"
the functions $\rho_j(x_0)=c_j$ and $\rho_0(x_0)=y_0$, associated
to $s$, we have immediately in particular
$\sg_0=s^\ast(\sgt_0)$, and
$\forall\,x_0$:
\begin{equation}
\sg_0\equiv d\rho_0-
\sum_{i=0}^{n}f_0^i(\rho)\,dx_0^i\,.
\label{sg0rho}
\end{equation}
We set
$\rho_{i}\equiv{}f_0^i(\rho)$.
Now, from (\ref{sg0rho}) and the pull-back of (\ref{dsg0}), one 
deduces that
\begin{equation}
\sum_{i=1}^{n}dx_0^i\wedge (d\rho_{i}-\sg_i)=0\,,
\label{exa}
\end{equation}
and in particular:
\begin{equation}
dx_0^1\wedge dx_0^2\wedge\dots\wedge dx_0^n\wedge
(d\rho_{i}-\sg_i)=0\,.
\end{equation}
Consequently
\begin{equation}
d\rho_{i}-\sg_i=\sum_{j=1}^{n}\rho_{i,j}\,dx_0^j
\Longleftrightarrow
\sg_i=d\rho_{i}-\sum_{j=1}^{n}\rho_{i,j}\,dx_0^j\,,
\end{equation}
that are alternatives to the pull-backs by $s$ of relations
(\ref{desy}). We thus have,
\begin{eqnarray}
&&\sg_0=
d\rho_0-\sum_{i=1}^{n}\rho_i\,dx_0^i\,,\label{sg0dr}\\
&&\sg_i=d\rho_i-\sum_{j=1}^{n}\rho_{i,j}\,dx_0^j\,.\label{sgidr}
\end{eqnarray}
Now, out of (\ref{sgidr}) and the pull-backs of
(\ref{dsgi}), one deduces also the relations,
\begin{equation}
\sum_{j=1}^{n}\,dx_0^j\wedge
d\rho_{i,j}=\sum_{j=1}^n\,dx_0^j\wedge\sg_{ij}\Longleftrightarrow
\sum_{j=1}^{n}\,dx_0^j\wedge
(d\rho_{i,j}-\sg_{ij})=0\,.
\end{equation}
By the same procedure as above, we thus get:
\begin{equation}
d\rho_{i,j}-\sg_{ij}=\sum_{j=1}^{n}\rho_{i,j,k}\,dx_0^k
\Longleftrightarrow
\sg_{ij}=d\rho_{i,j}-\sum_{k=1}^{n}\rho_{i,j,k}\,dx_0^k\,.
\end{equation}
Moreover, in view of (\ref{exa}) and (\ref{sgidr}) we deduce the
symmetries:
$\rho_{i,j}=\rho_{j,i}\equiv\rho_{ij}$ and
$\rho_{i,j,k}=\rho_{j,i,k}\equiv\rho_{ij,k}$. Then,
considering the coefficients of a same basis differential
form with $\boldsymbol{\rho}_2\equiv(\rho_{ij})$,
$\boldsymbol{\rho}_1\equiv(\rho_{i})$, and the system of
algebraic equations for  $\boldsymbol{\rho}^2_0$ deduced from
(\ref{dsg0}) and (\ref{dsgi}), we conclude that
$\boldsymbol{\rho}^2_0\in\Su$.
\hfill
$\Box$
\par\bigskip
In order to know the effects on
$\m$ of these infinitesimal deformations, we need to describe
what  are their incidences  upon the  objects acting
primarily on $\Rset^n$, namely the applications $\hf$. Thus,
we pass to the study of what we call the ``ab system" of the PDE system
(\ref{confeq}).
\par\medskip


\subsection{The ``ab system"} This system is defined
by the first two sets of PDE \eqref{f1} and \eqref{f2}. 
For this system of Lie equations, we will begin with recalling well-known
results but in the framework of the present context. Applying the same
reasoning than in the previous subsection,  we first
obtain the following results, which hold up to order two: 
\begin{subequations}
\begin{align}
&\sum_{r,s=1}^n\omega_{rs}(\hat{a}_0)
\,\,\hat{a}^{r}_{i}\,\hat{a}^{s}_{j}=
e^{2c_0}\omega_{ij}(x_0)\,,
\label{a1h}\\ 
&\hat{a}^k_{ij}+
\sum_{r,s=1}^n{}\gamma^{k}_{rs}(
\hat{a}_0)\,\,\hat{a}^{r}_{i}\,\hat{a}^{s}_{j}=
\sum_{q=1}^n{}\hat{a}_q^k\left(
\gamma^{q}_{ij}(x_0)+c_i\delta^q_j+
c_j\delta^q_i-\omega_{ij}(x_0)c^q\right)\,,
\label{a2h}
\end{align}
\label{a12h}
\end{subequations}
which clearly show that $J_1(\Rset^n)$ is
diffeomorphic to an embeded submanifold of the 2-jets
affine bundle $J_2(\Rset^n)$ of the $C^\infty(\Rset^n,\Rset^n)$
differentiable applications on
$\Rset^n$. In second place, we get relations, from the (T) system, for
the coefficients of order 3 that we only write as
($\hat{a}_1\equiv(\hat{a}^i_j)$;
$\hat{a}_2\equiv(\hat{a}^i_{jk})$,\dots,
$\hat{a}_k\equiv(\hat{a}^i_{j_1\cdots j_k})$;
$\bfah^k_0\equiv(\hat{a}_0,\cdots,\hat{a}_k)$):
\begin{equation}
\hat{a}^i_{jkh}\equiv
\hat{A}^i_{jkh}(x_0,\bfah^2_0)\,,
\label{a3h}
\end{equation}
where $\hat{A}^i_{jkh}$ are algebraic functions,
pointing out in this expression the independency from
the ``$c$" coefficients (also like the relations \eqref{a1h}, for
instance, when expressing $c_0$ with respect to the determinant of
$\hat{a}_1$). We denote by
$\Oc^i_J$ the Pfaff 1-forms at $x_0$ and $\{\hat{a}\}$
(or at  $(x_0,\{\hat{a}\})$):
\begin{equation}
\Oc^i_J\equiv
d\hat{a}^i_J-\sum_{k=1}^n\hat{a}^i_{J+1_k}dx^k_0\,,
\label{Oma}
\end{equation}
and  setting the $\hat{a}$'s as values of
functions $\hat{\tau}$ depending on $x_0$ (in some way we
make a pull-back on $\Rset^n$), we define the tensors $\kc$ by:
\begin{equation}
\Oc^i_J\equiv\sum_{k=1}^n\Big(\partial_k\hat{\tau}^i_J-
\hat{\tau}^i_{J+1_k}\Big)\,dx^k_0\equiv
\sum_{k=1}^n\kc^i_{k,J}\,
dx^k_0\,.
\label{omtau}
\end{equation}
Then from the relations:
\[
\e^{2c_0}\oms{r}{s}(\hat{a}_0)=
\sum_{i,j=1}^n\oms{i}{j}(x_0)\hat{a}^r_i
\hat{a}^s_j\,,\qquad
\sum_{i=1}^n\gam{i}{i}{k}=
\frac{1}{2}\sum_{i,j=1}^n\oms{i}{j}\,\partial_k\omi{i}{j}\,,
\]
we deduce with $\hat{b}\equiv\hat{a}_1^{-1}$, for example, that the
$\Oc^i_j$ 1-forms satisfy the relations at $(x_0,\bfah^1_0)$:
\begin{equation}
\widehat{H}_0(x_0,\bfah^1_0,\Oc^k_L;|L|\leq1)
\equiv\sum_{i,j=1}^n\hat{b}^j_i\,\Oc^i_j+
\sum_{j,k=1}^n\gamma^j_{jk}(\hat{a}_0)\Oc^k
=n\sg_0\,.\label{nsg0}
\end{equation}
Similar computations show that the 1-forms $\sg_i$
can be expressed as quite long relations, linear in the
$\Oc^j_J$ ($|J|\leq2$), with coefficients which are algebraic functions
depending on the $\hat{a}_K$ ($|K|\leq2$), the derivatives of the metric
and the Riemann-Christoffel symbols, all of them taken 
either at $x_0$ or $\hat{a}_0$. Then, we set:
\begin{equation}
\sg_i\equiv\widehat{H}_i(x_0,\bfah^2_0,\Oc^j_I;
|I|\leq2)\,.
\label{nsgi}
\end{equation}
From \eqref{a3h} the 1-forms 
$\Oc^i_{jkh}$  are also sums of 1-forms
$\Oc^r_K$ ($|K|\leq2$) with the same kind of coefficients
and not depending on the $\sg$'s, and we write:
\begin{equation}
\Oc^i_{jkh}\equiv\widehat{K}^i_{jkh}(x_0,
\bfah^2_0,\Oc^r_K;|K|\leq2)\,,
\label{K3}
\end{equation}
where $\widehat{K}^i_{jkh}$ are functions which are linear in the 1-forms
$\Oc^r_K$.\par\medskip
Let us denote by $\widehat{\text{\eu P}}_2\subset J_2(\Rset^n)$
the set of elements $(x_0,\bfah_0^2)$ satisfying relations
\eqref{a12h} whatever are the $c$'s.  Then the Pfaff system we
denote
$\widehat{P}_2$ over $\widehat{\text{\eu P}}_2$ and generated
by the 1-forms $\Oc^j_K\in{T^\ast\Rset^n}
\otimes_\Rset{J_2(\Rset^n)}$
in \eqref{Oma} with $|K|\leq2$, is locally integrable on
every neighborhood
$U{(x_0,\bfah_0^2)}{\subset}J_2(\Rset^n)$, since at $(x_0,\bfah^2_0)$
we have ($|J|\leq2$):
\begin{equation}
d\Oc^i_J-\sum_{k=1}^n dx^k_0\wedge\Oc^i_{J+1_k}\equiv0\,,
\label{dom}
\end{equation}
together with \eqref{K3}.\par\medskip
From now on, we consider the ``Poincar\'e system" whose 
corresponding notations will be free of ``hats". We denote by
$\Omega^i_J$ the Pfaff 1-forms corresponding to this
system, i.e. the system defined by the PDE
\eqref{f1} and \eqref{f2} with a vanishing function
$\alpha$. The corresponding 1-forms ``$\sg$"  are also
vanishing everywhere on $\Rset^n$ and the $\Omega^i_J$
satisfy all of the previous relations  with the $\sg$'s
cancelled out. Then it is easy to
see the 
$\Omega^i_J$ 1-forms ($|J|\geq2$) are generated
by the set of 1-forms $\Omega^j_K$ ($|K|\leq1$),
and in particular we have 
\begin{equation}
\Omega^k_{ij}=-\left\{\sum_{r,s,h=1}^n(\partial_h
\gamma^k_{rs})({a}_0)\Omega^h\,{a}^r_i\,{a}^s_j+\sum_{r,s=1}^n\gamma^k_{rs}({a}_0)[\,{a}^r_i\,\Omega^s_j+
{a}^s_j\,\Omega^r_i\,]\right\}\,,
\label{K2}
\end{equation}
with $(x_0,\bfah_0^1\equiv \bfa_0^1)\in\text{\eu P}_1\subset
J_1(\Rset^n)$, and $\text{\eu P}_1$ being the set of elements
satisfying relations \eqref{a1h} with $c_0=0$. Similarily
the Pfaff system we denote ${P}_1$ over $\text{\eu P}_1$
and generated by the 1-forms
$\Omega^j_K$ in \eqref{Oma} with $|K|\leq1$, is
locally integrable on every neighborhood
$U{(x_0,\bfa_0^1)}\subset\text{\eu P}_1$, since at the point
$(x_0,\bfa_0^1)$ we have  relations
\eqref{dom} with $|J|\leq1$ together with  relations
\eqref{K2}.\par
Then  at each  $(x_0,\bfah_0^2)\in\text{\eu P}_2$, we have
the locally exact splitted sequence
\begin{equation}
\begin{CD}
0@>>>P_1@>b_1>>\widehat{P}_2@>e_1>>P_c@>>>0\,,\label{pseq}
\end{CD}
\end{equation}
where we consider $J_1(\Rset^n)$ embeded in $J_2(\Rset^n)$ as well
as  $\text{\eu P}_1$ in $\widehat{\text{\eu
P}}_2\supset\text{\eu P}_2$ from relations \eqref{a12h}. In
this sequence a back connection
$b_1$  and a connection
$c_1:P_c\longrightarrow\widehat{P}_2$ are such that
($|J|\leq2$):
\begin{equation}
\Oc^i_J=\Omega^i_J+\chi^i_J(x_0,\bfah^2_0)\,\sg_0+
\sum_{k=1}^n\chi^{i,k}_J(x_0,\bfah^2_0)\,\sg_k\,,
\label{splitom}
\end{equation}
with $\Omega^i_{jk}$ satisfying \eqref{K2} for any given 
$\Omega^h_L$ with $|L|\leq1$, and where the tensors $\chi$ are defined on
$\text{\eu P}_2$. Together, they define  a back connection, and  the
tensors
$\chi$ define a
connection if they satisfy the relations:
\begin{align*}
&\widehat{H}_0(x_0,\bfah^1_0,\chi^k_L;|L|\leq1)=n\,,\qquad
\widehat{H}_0(x_0,\bfah^1_0,\chi^{k,i}_L;|L|\leq1)=0\,,\\
&\widehat{H}_i(x_0,\bfah^2_0,\chi^k_L;|L|\leq2)=0\,,\qquad
\widehat{H}_i(x_0,\bfah^2_0,\chi^{k,h}_L;|L|\leq2)=
n\delta^h_i\,,
\end{align*}
in order to preserve relations \eqref{nsg0} and 
\eqref{nsgi},  i.e. $e_1\circ c_1=id$.


\section{The  spacetime $\m$ unfolded by 
gravitation and electromagnetism} Again, in view of physical
interpretations, we put a spotlight on the tensor $\B$. In fact, we
consider the relations \eqref{splitom} with $|J|=0$ and
the $\Oc^i$ as fields of {\em``tetrads"\/}. Then we get a  metric
$\nu$ for the {\it``unfolded spacetime $\m$"\/}
defined infinitesimally at $x_0$ (corresponding to $p_0$ in $\m$) by:
\begin{align*}
&\nu=\tilde{\omega}=\omega+\delta\omega=
\sum_{i,j=1}^n\omega_{ij}\circ\hat\tau(x_0)\,
\Oc^i(x_0)\otimes\Oc^j(x_0)\,,\\
&\Oc^i=\sum_{k=1}^n\kc^i_{k}(x_0)\,
dx^k_0\,,\\
&\kc^i_j(x_0)=\kappa^i_j(x_0)+\chi^i(x_0,\bftau_0^2)\,\A_j(x_0)+
\sum_{k=1}^n\chi^{i,k}(x_0,\bftau_0^2)\,\B_{j,k}(x_0)\,.
\end{align*}
We consider the particular case for which the S-admissible metric
$\omega$ is equal to\linebreak $\text{diag}[+1,-1,\cdots,-1]$ (and thus $k_0=0$),
the $\chi$'s are only depending on $x_0$  and
$\kappa^i_j=\delta^i_j$,  i.e. the deformation of
$\omega$ is only due to the tensors $\A$ and
$\B$. Thus, one has the general relation between
$\nu$ and $\omega$: $\nu=\omega\,+$ linear and quadratic
terms in $\A$ and $\B$. Then from this metric
$\nu$, one can deduce the Riemann and Weyl curvature
tensors of the {\it``unfolded spacetime $\m$"\/}.\par 
We are faced with a question: could this kind of deformation
be interpreted as an inflation  process in cosmology (due to a kind of
instable substratum spacetime $\Sub$ for instance)   ?  In such a
process, each occurrence  of a  creation or annihilation of
singularities  of
the gauge potentials $\A$ and $\B$, would be
related to a non-trivial unfolding, i.e. a non-smooth deformation.\par
Also, would the inflation be the evolution from the Poincar\'e Lie
structure to ``a" conformal one, and going from a physically
``homogeneous" spacetime (namely with constant Riemann curvature and a
vanishing Weyl tensor and so {\it ``conformally flat"\/}) to an ``inhomogeneous" one
(with any Weyl tensor)~? Or, would the substratum spacetime $\Sub$
be merely a kind of ``mean" smooth manifold $\m$ with no inflation
process ?\par In some way, the singularities of the gauge potentials
$\A$ and $\B$ would produce a kind of {\it ``bifurcation"\/} of the
unfolded spacetime structure leading to a different concept of
bifurcation than the one used in the case of non-linear
ODE's.\par\medskip  In view of making easier computations for a
relativistic action deduced from the metric tensor
$\nu$, we consider this metric in the {\it ``weak fields limit"\/},
assuming that the metric $\nu$ is linear in the tensors $\A$  and $\B$
and that the quadratic terms can be neglected. Furthermore, from 
relations
\eqref{ABIJeq}, we have  the
relations:
\[\partial_i\A_k-\partial_k\A_i=\B_{k,i}-\B_{i,k}=
\F_{ik}\,,
\qquad
\partial_j\B_{k,i}-\partial_k\B_{j,i}\simeq
0\,,
\]
since  the functions $\rho$ take also small
values in the weak fields limit. Therefore, we can write 
\(
\nu_{ij}\simeq\omega_{ij}+\epsilon_{ij}\,
\),
where the  coefficients $\epsilon_{ij}$ can be viewed as small
perturbations of the metric $\omega$ and linearly defined from
$\A$, $\B$ and the $\chi$ tensors.
Then, let $i$ be a differential map
$i:s\in[0,\ell]\subset\Rset\longrightarrow
i(s)=x_0\in\Rset^n\simeq\m$, and $U$ being such that 
$U\equiv di(s)/ds\,$,
${\parallel}U{\parallel}^2=1$. We define the
relativistic action
$S_1$ by:
\begin{equation*}
S_1=\int_0^\ell\sqrt{\nu(U(s),U(s))}\;ds
\equiv\int_0^\ell \sqrt{L_\nu}\;ds\,.
\end{equation*}
We also take the tensors
$\chi$ as depending on $s$ only. The Euler-Lagrange equations
for the Lagrangian density $\sqrt{L_\nu}$ are not
independent because $\sqrt{L_\nu}$ is a homogeneous
function of degree 1 and thus satisfies an additional
homogeneous differential equation. Then, it is well-known
that the variational problem for
$S_1$ is equivalent to consider the variation of the  
action $S_2$ defined by
\begin{equation*}
S_2=\int_0^\ell{\nu(U(s),U(s))}\;ds
\equiv\int_0^\ell L_\nu\;ds\,,
\end{equation*}
but constrained by the condition $L_\nu=1$. In this
case,  this shows that $L_\nu$ must be
considered, firstly,  with an associated Lagrange multiplier, namely a
mass, and secondly, its explicit expression with respect to $U$
will appear only in the variational calculus. In the weak
fields limit, we obtain:
\begin{equation}
L_\nu={\parallel}U{\parallel}^2+
2\sum_{j,k=1}^n\omega_{kj}\,\chi^k\,U^j\;.\;
\sum_{i=1}^n\A_i\,U^i
+2\sum_{j,k,h=1}^n\chi^{k,h}\,\omega_{kj}\,U^j\;.\;
\sum_{i=1}^nU^i\,\B_{i,h}\,.
\label{lnu}
\end{equation}
From the latter relation, we can deduce a few
physical consequences among others. On the one hand, if we assume
that
\begin{subequations}
\begin{align}
\mathcal{C}^h(\chi,U)\equiv&
\sum_{j,k=1}^n\omega_{kj}\chi^{k,h}U^j\equiv\zeta^h\,,
\label{thomgen}\\
\mathcal{C}^0(\chi,U)\equiv&
\sum_{k,j=1}^n\omega_{kj}\chi^kU^j\equiv\zeta_0\,,
\label{thomas}
\end{align}
\label{prec}
\end{subequations}
then we recover in
\eqref{lnu}, up to some suitable constants, the
Lagrangian density for a particle, with the {\it
velocity n-vector $U$\/}
(\({\parallel}U{\parallel}^2=1\)), embeded in an
external electromagnetic field. But also from the relation
\eqref{thomas} we find {\it``a generalized Thomas precession"\/}
if  the tensor $(\chi^k)$ is ascribed (up to a suitable constant for units) to
a {\it``polarization
$n$-vector"\/} \cite[p.~270]{bacry} {\it``dressing"\/} the
particle (e.g. the spin of an electron). This
generalized precession could give a possible origin for
the creation of anyons in high-$T_c$ superconductors
\cite{rub94} and might be an alternative to
Chern-Simon theory. Also,  the tensor
$(\chi^{k,h})$ might be a \textit{polarization tensor\/} of some matter 
and the particle would be ``dressed" with this kind of
polarization.\par\medskip
More generally, the Euler-Lagrange equations associated to
$S_2$ would define a system of geodesic equations  with
Riemann-Christoffel symbols $\Gamma$ associated to $\nu$ and such
that (with \(\nu^{ij}\simeq\omega^{ij}\) at first order 
and assuming the $\chi$'s being
constants)
\begin{equation}
\frac{\;\,dU^r}{ds}=-\sum^n_{j,k=1}
\Gamma_{jk}^r\;U^j\,U^k
+\xi_0\sum_{i,k=1}^n\omega^{kr}\,{\F_{ki}}\,U^i\,,
\label{geod}
\end{equation}
with
\begin{equation*}
\Gamma_{jk}^r=\frac{1}{2}\left(\chi^r(\partial_k\A_j+\partial_j\A_k)+
\sum_{h=1}^n\chi^{r,h}
\left(\partial_j\B_{k,h}+\partial_k\B_{j,h}\right)\right)\,,
\end{equation*}
and $\A$ and $\B$ satisfying the first and second sets of
differential equations (compare \eqref{geod} with the analogous equation
($6.9''$) in \cite{fulton62} but with different Riemann-Christoffel
symbols). Then the tensor
$\Gamma$ would be associated to gravitational fields, also providing
other physical interpretations for the tensors $\chi$.\par
Nevertheless,  equations \eqref{geod} are deduced
irrespective of the conditions \eqref{prec}. Taking them into account 
would lead to a modification of the action $S_2$ resulting of the
introduction of  Lagrange multipliers $\lambda_0$ and $\lambda_k$
($k=1,\cdots,n$) in the Lagrangian density definition. We would then
define a new action:
\begin{equation*}
{S}_2=\int_0^{\ell}
\left\{m{\parallel}U{\parallel}^2+\sum_{i=1}^n
\epsilon_{i}\,U^i
-\sum_{k=0}^n\lambda_k\,\mathcal{C}^k(\chi,U)
\right\}ds\,.
\end{equation*}
The associated Euler-Lagrange equations would be analogous to
\eqref{geod}, but with additional
terms coming from the precession. Moreover, since we have the
constraint
${\parallel}U{\parallel}^2=1$, we need a new Lagrangian
multiplier denoted by $m$. That also means we do
computations on the projective spaces $H(1,n-1)$.
From this  point of view, the Lagrange multipliers appear to
be non-homogeneous coordinates of these projective spaces.\par
Then the variational calculus
would also lead to additional precession equations giving
torsion as mentioned in  comments of chapter 1. Again, torsion is
not related to  unification but  to parallel
transports on manifolds which is a well-known geometrical
fact \cite{dieudon}. Hence, the existence of such
precession phenomenon for a spin $n$-vector $(\chi^k)$ would be
correlated with the existence of linear ODE for a charged particle of
charge $\xi_0$ interacting with an electromagnetic field. Otherwise
without \eqref{thomas} the ODE's would be non-linear and there wouldn't
be  any kind of precession  of any spin \mbox{$n$-vector}.\par
Consequently the  motion defined by the second term in the r.h.s. of
\eqref{geod} for a spinning charged particle  would just be, in this
model,  a point of view  resulting from an implicit separation of
rotations and translations degrees of freedom achieved by the
specialized (sensitive to particular subgroups of the  symmetry group of
motions) experimental apparatus in
$T_{p_0}\m$. The latter separation would insure either some simplicity
i.e. linearity, or, since the measurements are achieved in
$T_{p_0}\m$, that the equations of motion are associated (\textit{via\/}
some kind of projections inherent to implicit dynamical constraints due
to the experimental measurement process, fixing,  for instance,
$\xi_0$ to a constant) to linear representations of 
tangent actions of the Lorentz Lie group on $T_{p_0}\m$. In the latter
case, one could say, a somewhat provocative way, that special relativity
covariance would have to be satisfied,
\textit{as much as possible\/}, by physical laws. This ``reduction" to
linearity can't be done on the first summation in the r.h.s. of
\eqref{geod}, which must be left quadratic contrarily to the second one,
since the Riemann-Christoffel symbols can't be defined equivariantly
(other arguments can be found in~\cite{fulton62}).\par To conclude,
equations
\eqref{geod} would provide us with another interpretation of spin (the
$\chi$'s) as an object allowing moving particles to generate effective 
spacetime deformations, as ``wakes" for instance.


\section{Conclusion: images set back from their objects}
In fact in this work, using the Pfaff systems theory and the Spencer
theory of differential equations, we studied the formal solutions of the
conformal Lie system with respect to the Poincar\'e one. More precisely,
we determined the difference between these two sets of formal solutions.
We gave a description of a ``relative" set of PDE, namely the ``c
system" which defines a deployment from the Poincar{\'e} Lie pseudogroup
to a  sub-pseudogroup of the conformal Lie pseudogroup.  We studied
these two systems of Lie equations because of their occurrence in
physics,  particularly in electromagnetism as well as in Einsteinian
relativity.\par On the basis of this concept of deployment, we made the
assumption that the unfolding is related to the existence of two kind of
spacetimes, namely: the substratum or striated spacetime $\Sub$ from
which the 4D-ocean or smooth spacetime $\m$ is unfolded. We recall that
not all the given metrics on $\Sub$ can be admissible in order to have
such spacetime $\m$. In the case of a substratum spacetime $\Sub$
endowed with an appropriate S-admissible metric allowing for unfolding,
we assumed that $\Sub$ is  equivariant with respect to  the conformal 
and  Poincar\'e pseudogroups and  set its Riemannian scalar curvature 
to  a constant $n(n-1)k_0$ and its Weyl tensor to zero. Then the deployment evolution can be trivial
or not depending on occurences of spacetime singularities (of the
potentials of interaction forces) parametrizing or dating what can be
considered somehow as a kind of  spacetime history. The  potentials of
interactions are built out of  a particular relative Spencer 
differential sequence associated to the ``c system" and describing
smooth deformations of $\Sub$. Then a ``local" metric or ``ship-metric"
defined on a moving tangent spacetime  ship $\tom$ of the unfolded
spacetime $\m$ is constructed out of the S-admissible substratum
metric and of the deformation potentials. The spacetime ships dynamics 
are given by a system of PDE satisfied by their  Lorentzian  velocity
$n$-vectors $U$,  exhibiting both classical electrodynamic and
``local" (since ship-metrics are local) geodesic navigation in spacetime
endowed with gravitation.\par This is  a ``classical approach'' and
quantization doesn't seem to appear. Nevertheless, in the present
framework, a few physical exotic quantum
effects and experiments might be revisited from a classical
viewpoint: for instance the Aharanov-B\"ohm effect and the ``one plate
dynamical Casimir effect". Indeed, the first summation in the r.h.s. of
\eqref{geod} could give a classical effect similar to the
Aharanov-B\"ohm quantum one \cite{aharanovbohm} and could be
another origin for the creation of anyons in superconductors. Usually the
symmetric part of the 4-gradient of
$\A$ appearing in the expression for $\Gamma$ is  never taken into
account in classical physics. But, and it is a deep problem, it is
contradictory with quantum mechanics since in that case the 4-vector
$\A$  (and not its 4-gradient) is involved in the Shr\"odinger equation
for electromagnetic interactions. In other words, $\A$ defines a set of
physical observables and, as a consequence, the symmetric part of its
4-gradient would have to define also other physical observables. These
latter never appear in classical electrodynamics and in  usual
classical physics. In some way, the occurence of this symmetric part in
$\Gamma$ is much more coherent with  quantum mechanics and would not
be meaningless.\par Also, going on again in a non-rigorous metaphoric
description, could the one plate (linear) dynamical Casimir effect 
\cite{casimirmoore,fullingdavies} be analogous to the
\textit{cavitation\/} of a paddle or of a fan blade of a
screw-propeller in a spacetime 4D-ocean, with steam bubbles being made
of photons and/or gravitons~? This cavitation effect would occur as early
as the motion of the uncharged metallic plate wouldn't be  an uniformly
accelerated one. Then the conformal spacetime symmetry would be broken
and  occurence  of bubbles of ``electromagnetic and gravitational steam"
dressing the plate  would restore only the conformal symmetry when
considering the full resulting system. In the same vein, gravitational
waves would be radiated only by non-uniformly accelerated massive
bodies. Also, on the basis of the previous Casimir effect with its
screw-propeller metaphor, the case of a non-uniformly accelerated
uncharged conductive rotating  disk would have to be   experimentally
considered in order to investigate new forms and conditions of
electromagnetic and/or gravitational radiations. This would be a one
plate \textit{rotational\/} dynamical Casimir effect.\par\medskip To
finish on a more philosophical note, which is rather unavoidable since
we are concerned with  spacetime structure, the model we present is
strongly related to the H. Bergson one in his well-known book about the
A. Einstein relativity: ``Dur{\'e}e et Simultan{\'e}it{\'e}"
\cite{bergsondur}. It seems that H. Bergson was  misunderstood
and   opposed to  Einstein relativity whereas exactly the converse
appears to be true.  In fact H.
Bergson claimed 1) that A. Einstein relativity is a confirmation of the
existence of a universal time (i.e. history and not duration) because of
an implicit reciprocity involved in the relativity principles and 2)
that A. Einstein gave the keys to interpret some ``appearances"
resulting from observations between \textit{Galilean frames\/} (which involves to
consider the Poincar{\'e} pseudogroup only). As a
particular result the twin paradox of M. Langevin doesn't exist any more
as H. Bergson shown in appendix III of
\cite{bergsondur}. In the framework of the model we present, this can be
easily shown by making integrations along universe lines by using the results
at the end of the chapter 3 about the transformation laws between Galilean
frames with velocity $n$-vectors $\dot{x}$ and
$\dot{\hx}^i=\sum_{j=1}^n\hf^i_j\,\dot{x}^j$. The ``time"  integrals,
i.e. the actions $S_1$, are coordinate (topological) invariant and
equal. In other words, following a kind of G. Berkeley philosophy of
eyesight strongly related to our Quattrocento painters  metaphor evoked
in the first chapter, the ``physical reality" would have to  be
interpreted with the help of the Lorentz transformations. Indeed special
relativity measurements would be achieved somehow only on ``observed
images set back from their  objects and following them like their
shadows", and as far away from their objects and slowly varying as the
relative velocity between their objects and the frames of measurements
increase. This is a Doppler-Fizeau effect.\par We can also point out some 
recent reflexions of\, G. 't Hooft about obstacles on the way towards 
the quantization of space, time and matter \cite{thooftphil}. 
His theory of ``ontological states"
involved in his approach of ``deterministic quantization", strongly
requires such ``universal time" and a description of spacetime as a fluid
(his own words) as well as its relationship with the principle of
coordinate invariance. A very analogous approach of the 't Hooft one for
the time irreversibility,  can be considered with the Prigogine concept
of thermodynamical time operator, in the framework of the Kolmogorov
flows for PDE~\cite{misraK}. It would be applied to the non-linear
``c~system". Thus, there would be two kind of times: a unique global
cosmic time (as in chapter 1) orienting local thermodynamical ones, the
same way as the terrestrial magnetic field orients the magnetic needles
of compasses.
\bibliography{Ether_Pfaff}
\bibliographystyle{plain}

\end{document}